\documentclass[longauth,bibyear]{aa}
\def\MCN{\mbox{CH$_3$CN}}

\def\MCNII{CH${_3}^{13}$CN}

\def\HII{H{\sc ii} }

\def\UC{UC~H{\sc ii}}

\def\mjy{~mJy~beam$^{-1}$}

\def\g31{G31}

\newcommand{\be}{\begin{equation}}
\newcommand{\ee}{\end{equation}}

\begin{document}
\title{Fragmentation in the massive G31.41+0.31 protocluster\thanks{The reduced images (FITS files) are only available at the CDS via
anonymous ftp to cdsarc.u-strasbg.fr (130.79.128.5) or via
http://cdsarc.u-strasbg.fr/viz-bin/qcat?}}
%\title{The sharp ALMA and VLA views of the high-mass core G31.41+0.31} 
\author{M.\ T.\ Beltr\'an\inst{1}, V.\ M.\ Rivilla\inst{2, 1}, R.\ Cesaroni\inst{1}, L.\ T.\ Maud\inst{3}, D.\ Galli\inst{1}, L.\ Moscadelli\inst{1}, A.\ Lorenzani\inst{1}, A.\ Ahmadi\inst{4}, H.\ Beuther\inst{5}, T.\ Csengeri\inst{6}, S.\ Etoka\inst{7}, C.\ Goddi\inst{8}, P.\ D.\ Klaassen\inst{9}, R.\ Kuiper\inst{10}, M.\ S.\ N.\ Kumar\inst{11}, T.\ Peters\inst{12}, \'A.\ S\'anchez-Monge\inst{13}, P.\ Schilke\inst{13}, F. van der Tak\inst{14, 15}, S.\ Vig\inst{16}, H.\ Zinnecker\inst{17}
}
\institute{
INAF-Osservatorio Astrofisico di Arcetri, Largo E.\ Fermi 5,
I-50125 Firenze, Italy
\and
Centro de Astrobiolog\'{i}a (CSIC-INTA), Ctra.\ de Ajalvir Km. 4, Torrej\'on de Ardoz, 28850 Madrid, Spain
\and
European Southern Observatory, Karl-Schwarzschild-Str.\ 2, 85748 Garching, Germany
\and
Leiden Observatory, Leiden University, PO Box 9513, 2300 RA Leiden, The Netherlands 
\and
 Max-Planck-Institut f\"ur Astronomie, K\"onigstuhl 17, 69117 Heidelberg, Germany
\and
Laboratoire d’Astrophysique de Bordeaux, Univ.\ Bordeaux, CNRS, B18N, all\'ee Geoffroy Saint-Hilaire, 33615 Pessac, France 
\and
Jodrell Bank Centre for Astrophysics, The University of Manchester, Alan Turing Building, Manchester M13 9PL, UK
\and
Department of Astrophysics, Institute for Mathematics, Astrophysics and Particle Physics (IMAPP), Radboud University, P.O.\ Box 9010, 6500 GL Nijmegen, The Netherlands
\and
UK Astronomy Technology Centre, Royal Observatory Edinburgh, Blackford Hill, Edinburgh EH9 3HJ, UK
\and
Institute of Astronomy and Astrophysics, University of T\"ubingen, Auf der Morgenstelle 10, 72076 T\"ubingen, Germany
\and
Instituto de Astrof\'{i}sica e Ci\^encias do Espa\c{c}o, Universidade do Porto, CAUP, Rua das Estrelas, 4150-762, Porto, Portugal
\and
Max-Planck-Institut f\"{u}r Astrophysik, Karl-Schwarzschild-Str. 1, D-85748 Garching, Germany
\and
I.\ Physikalisches Institut, Universit\"at zu K\"oln, Z\"ulpicher Str. 77, 50937 K\"oln, Germany
\and
Kapteyn Astronomical Institute, University of Groningen, 9700 AV Groningen, The Netherlands
\and
SRON Netherlands Institute for Space Research, Landleven 12, 9747 AD Groningen, The Netherlands
\and
Department of Earth and Space science, Indian Institute of Space Science and Technology, Thiruvananthapuram, 695 547, India
\and
Universidad Aut\'onoma de Chile, Avda. Pedro de Valdivia 425, Providencia, Santiago de Chile, Chile
}
\offprints{M.\ T.\ Beltr\'an, \email{maria.beltran@inaf.it}}

\date{Received date; accepted date}

\titlerunning{Fragmentation in G31.41+0.31}
\authorrunning{Beltr\'an et al.}

\abstract
{ALMA observations at 1.4\,mm and $\sim$0$\farcs$2 ($\sim$750\,au) angular resolution of the {\it Main} core in the high-mass star forming region G31.41+0.31 have revealed a puzzling scenario: on the one hand, the continuum emission looks very homogeneous and the core appears to undergo solid-body rotation, suggesting a monolithic core stabilized by the magnetic field; on the other hand, rotation and infall speed up toward the core center, where two massive embedded free-free continuum sources have been detected, pointing to an unstable core having undergone fragmentation.}
{To establish whether the {\it Main} core is indeed  monolithic or its homogeneous  appearance is due to a combination of large dust opacity and low angular resolution, we carried out millimeter observations at higher angular resolution and different wavelengths.}
{We carried out ALMA observations at 1.4\,mm and 3.5\,mm that achieved angular resolutions of  $\sim$0$\farcs$1 ($\sim$375\,au) and $\sim$0$\farcs$075 ($\sim$280\,au), respectively. VLA observations at 7\,mm and 1.3\,cm at even higher angular resolution, $\sim$0$\farcs$05 ($\sim$190\,au) and $\sim$0$\farcs$07 ($\sim$260\,au), respectively, were also carried out to better study the nature of the free-free continuum sources detected in the core.}
{The millimeter continuum emission  of the {\it Main} core has been clearly resolved into at least four sources, A, B, C, and D, within 1$''$, indicating that the core is not monolithic. The deconvolved radii of the dust emission of the sources, estimated at 3.5\,mm, are  $\sim$400--500\,au, their masses range from $\sim$15 to $\sim$26\,$M_\odot$, and their number densities are  several 10$^9$\,cm$^{-3}$. Sources A and B, located closer to the center of the core and separated by $\sim$750\,au, are clearly associated with two free-free continuum sources, likely thermal radio jets, and are the brightest in the core. The spectral energy distribution of these two sources and their masses and sizes are similar and suggest a common origin. Source C has not been detected at centimeter wavelengths, while source D has been clearly detected at 1.3\,cm. The fact that source D is likely the driving source of an E--W SiO outflow previously detected in the region suggests that the free-free emission is possibly arising from a radio jet.} 
{The observations have confirmed that the {\it Main} core in G31 is collapsing, has undergone fragmentation and that its homogeneous appearance previously observed at short wavelengths is a consequence of both large dust opacity and insufficient angular resolution. The low level of fragmentation together with the fact that the core is moderately magnetically supercritical, suggests that G31 could have undergone a phase of magnetically-regulated evolution characterized by a reduced fragmentation efficiency, eventually leading to the formation of a small number of relatively massive dense cores.}

\maketitle

\section{Introduction}
\label{sect-intro}

Despite the many theoretical and observational efforts aimed at understanding the formation process of early-type stars, only limited progress has been made. While there is growing evidence that B-type or late O-type stars, with masses up to $\sim$20$\,M_\odot$, may form through disk accretion (Beltr\'an \& de Wit~\cite{beltran16}), the number of early O-type (proto)stars ($L\ga 10^5~L_\odot$) with Keplerian rotation signatures is still very limited (see Beltr\'an~\cite{beltran20}, and references therein). Therefore, it is not yet clear if disk accretion is their only/main formation mechanism. These objects are rare, distant, and lie in complex regions inside rich stellar clusters. All these features make it difficult from an observational point of view to characterize the environment around the star and thus compare theoretical predictions (e.g., Bonnell \& Bate~\cite{bonnell06}; Krumholz et al.~\cite{krumholz09}; Tan et al.~\cite{tan14}; Motte et al.~\cite{motte18}; Kuiper \& Hosokawa~\cite{kuiper18}; Kumar et al.~\cite{kumar20}). It is also possible that the complexity of such an environment (Peters et al.~\cite{peters10}) makes any model proposed so far too simple to be satisfactory.

In an attempt to shed light on this issue and, in particular, establish the role of circumstellar disks in the formation of O-type stars, Cesaroni et al.~(\cite{cesa17}) performed a mini-survey at 1\,mm with the Atacama Large Millimeter/submillimeter Array (ALMA) at an angular resolution of $\sim$0$\farcs$2 of six luminous ($\ga5\times10^4~L_\odot$) sources associated with known hot molecular cores (HMCs). These observations have revealed the presence of velocity gradients consistent with rotation in almost all cores and different degrees of fragmentation, with two cores presenting no evidence of fragmentation. 

One of the HMCs that showed no signs of fragmentation was G31.41+0.31 (hereafter G31). G31 is a well known HMC located at 3.75~kpc (estimated from trigonometric parallax; Immer et al.~\cite{immer19}), with a luminosity of $\sim 4.5\times10^4~L_\odot$ (from Osorio et al.~\cite{osorio09}, after scaling to the parallactic distance). 
The core is $\sim$5$''$ away from an ultracompact (UC) \HII region and harbors two point-like ($<$0\farcs07) free-free continuum sources close to the center separated by $\sim0\farcs2$ (Cesaroni et al.~\cite{cesa10}, hereafter C10).
Given the intensity of the molecular line and continuum emission, interferometric studies at millimeter wavelengths with sub-arcsecond resolution ($\sim$0\farcs8; Cesaroni et al.~\cite{cesa11}) have been
feasible even before the advent of ALMA. Such observations have confirmed that the core is chemically very rich, presenting prominent emission in a large number of complex organic molecules (Beltr\'an et al.~\cite{beltran05}, \cite{beltran09}; Fontani et al.~\cite{fontani07}; Calcutt et al.~\cite{calcutt14}; Rivilla et al.~\cite{rivilla17}; Mininni et al.~\cite{mininni20}). These have been successfully used to trace the distribution and kinematics of the dense hot gas, and to estimate basic physical parameters (e.g. temperature and density; Beltr\'an et al.~\cite{beltran05}). The presence of a clear velocity gradient along the NE--SW direction in the core has been established (Beltrán et al.~\cite{beltran04}; Girart et al.~\cite{girart09}; Cesaroni et al.~\cite{cesa11}). This gradient has been interpreted as rotational motion around embedded massive stars, whose existence is suggested by the presence of the two free-free continuum sources (C10). Polarization measurements by Girart et al.~(\cite{girart09}) and Beltr\'an et al.~(\cite{beltran19}) have revealed an hour-glass shaped magnetic field with the symmetry axis oriented perpendicular to the velocity gradient. Moreover, inverse P-Cygni profiles in a few molecular lines have been detected toward the core (Girart et al.~\cite{girart09}; Mayen-Gijon et al.~\cite{mayen14}), indicating that infall motions are present. All these features are consistent with a scenario where the core is collapsing and rotating about the direction of the magnetic field axis.

In a recent study, Beltr\'an et al.~(\cite{beltran18}, hereafter B18) analyzed in detail the line and continuum ALMA observations at 1\,mm with $\sim$0\farcs2 ($\sim$750~au) angular resolution initially presented in Cesaroni et al.~(\cite{cesa17}). This high angular resolution allowed us to clearly resolve the core for the first time, whose angular diameter is $\sim$2\arcsec. The most remarkable results of this study were the following:

\begin{itemize}

\item The dust continuum emission has been resolved into two cores, a {\it Main} core 
that peaks close to the position of the two free-free continuum sources detected by C10, and a much weaker and smaller {\it NE} core. 

\item The {\it Main} core imaged in the continuum is well resolved and looks monolithic and featureless, with no hint of fragmentation, despite the large mass ($\sim$70\,$M_\odot$; Cesaroni et al.~\cite{cesa19}) and the diameter of $\sim$8000\,au.
 
\item The velocity gradient along the NE--SW direction observed in the {\it Main} core is due to a smooth change of velocity across the core, which appears to increase linearly with  distance from the core center, consistent with solid-body rotation, as shown e.g., in the position-velocity (PV) plot in Fig.~8 of B18.
 
\item The rotation seems to speed up with increasing energy of the CH$_3$CN transition, with the PV plot in the vibrationally excited \MCN(12--11) $v_8$=1 $K,l$=6,1 ($E_{\rm up}$=778~K) line being steeper than the same plot for the ground-state \MCNII(12--11) $K$=2 ($E_{\rm up}$=97~K) line (see Fig.~9 of B18). This spin-up toward the center of the rotation is expected from conservation of angular momentum in a rotating and infalling structure.

\item The detection of red-shifted absorption toward the core center supports
 the presence of infall in the core. Interestingly, the velocity of the
 absorption feature increases with the line excitation energy, which suggests that infall is accelerating toward the core  center (where the temperature is higher), consistent with gravitational collapse on a central source.
\end{itemize}

\begin{table*}
\caption[] {Summary of the continuum observations$^a$.}
\label{table_parobs}
\begin{tabular}{cccccccc}
\hline
&&&&\multicolumn{2}{c}{Synthesized beam}  
\\
\cline{5-6}
&&&&\multicolumn{1}{c}{HPBW} &
\multicolumn{1}{c}{P.A.} &
\multicolumn{1}{c}{rms} 
\\
\multicolumn{1}{c}{Interferometer} &
\multicolumn{1}{c}{Wavelength} &
\multicolumn{1}{c}{Configuration} &
\multicolumn{1}{c}{Epoch} &
\multicolumn{1}{c}{(arcsec)} &
\multicolumn{1}{c}{(deg)} &  
\multicolumn{1}{c}{(\mjy)} 
\\
\hline
ALMA &1.4 mm &C40-7  &2017 Aug. 16, 17  &$0.114\times0.084^b$  &$-82^b$     &0.320$^b$ \\
&&C34-7/(6)	 &2015 Jul.\ 20, 21  \\
ALMA &3.5 mm &C40-6, C43-6,   &2016 Sep. 30,
2017 Sep. 17, &$0.077\times0.074^b$  &20$^b$     &0.160$^b$\\
&&C40-9, C43-9 &2019 Jun. 23, 2019 Aug. 20    \\
VLA &7 mm & A & 2018 Apr. 9, 22 &$0.051\times0.043$  &23     &0.025 \\
VLA &1.3 cm & A  &  2018 Mar. 25, Apr. 14, 15 &$0.076\times0.067$  &8     &0.005 \\
\hline
\end{tabular}

$a$ The phase reference center of the observations was set to the position $\alpha$(J2000) = 18$^{\rm h}$ 47$^{\rm m}$ 34$\fs$315, $\delta$(J2000) = $-$01$^\circ$ 12$'$ 45$\farcs$9. \\
$^b$ Synthesized beam and rms noise obtained after combining all the data.  
\end{table*}

These results altogether pose serious questions about the nature of the core. On the one hand, the presence of a magnetic field
uniformly threading the core (Girart et al.~\cite{girart09}; Beltr\'an et al.~\cite{beltran19}), the linear increase of rotation velocity with radius, and the (apparent) homogeneity of the core, point to a {\it monolithic core stabilized by the magnetic field and undergoing solid-body rotation}, similar to the scenario proposed by the  ``monolithic collapse'' theory (also known as turbulent core model) of McKee \& Tan~(\cite{mckee03}). On the other hand, the presence of red-shifted absorption, the existence of two embedded massive stars (the two free-free sources), and the rotational speed-up toward the core center are consistent with an {\it unstable core undergoing fragmentation with infall and differential rotation due to conservation of angular momentum}, a scenario reminiscent of the ``competitive accretion'' model of Bonnell \& Bate~(\cite{bonnell06}). However, the striking monolithic appearance of the {\it Main} core may be the result of a combination of high opacity of the 220~GHz continuum emission and insufficient angular resolution. In fact, we may reasonably assume that the size of a typical fragment is of the same order as the separation between the two free-free sources detected inside the core, $\sim$0\farcs2 or $\sim$750\,au, comparable to the synthesized beam of our ALMA maps, which could have prevented B18 from properly imaging the fragmentation in the core -- if any.  At the same time, we must take into account that the continuum emission observed at 220~GHz is optically thick, as proved by the high brightness temperature of the continuum peak ($\sim$130~K) compared to the dust temperature ($\ga$150~K: Beltr\'an et al.~\cite{beltran05}) and the presence of deep absorption against this bright continuum, seen in many lines (Rivilla et al.~\cite{rivilla17}; B18). The large opacity %($\la$2) 
is likely to mask any inhomogeneity of the core.

To discriminate between the two scenarios depicted above for the G31 core, either a (marginally) stable monolithic core undergoing solid-body rotation, or a dynamically
collapsing core undergoing global fragmentation, we observed the G31 core with ALMA  at higher angular resolution, i.e.\ $\sim$0$\farcs$1 (375~au), at 1.4\,mm,  which should allow us to resolve the single fragments and estimate their masses. In addition, we also observed the core at a slightly higher angular resolution ($\sim$0$\farcs$075 or $\sim$280\,au) at 3.5\,mm, where the dust continuum opacity may be up to $\sim$4 times lower than at 1.4\,mm. These observations should allow us to discard or confirm that the core is indeed monolithic as it appears in the previous 1.4\,mm observations. We also carried out Very Large Array (VLA) observations at 7\,mm and 1.3\,cm at an even higher angular resolution (0$\farcs$05--0$\farcs$07) to better study the nature of the free-free continuum sources detected by C10 and possibly detect other fainter sources embedded in the core.

This article is organized as follows: in Sect.~2 we describe the ALMA and VLA
observations; in Sect.~3 we present the results and analyze the continuum emission at millimeter and centimeter wavelengths toward G31; in Sect.~4 we discuss the most likely scenario for G31, and the role of the magnetic field.  Finally, in Sect.~5  we give our main conclusions.

\begin{figure*}
\centerline{\includegraphics[angle=0,width=17cm,angle=0]{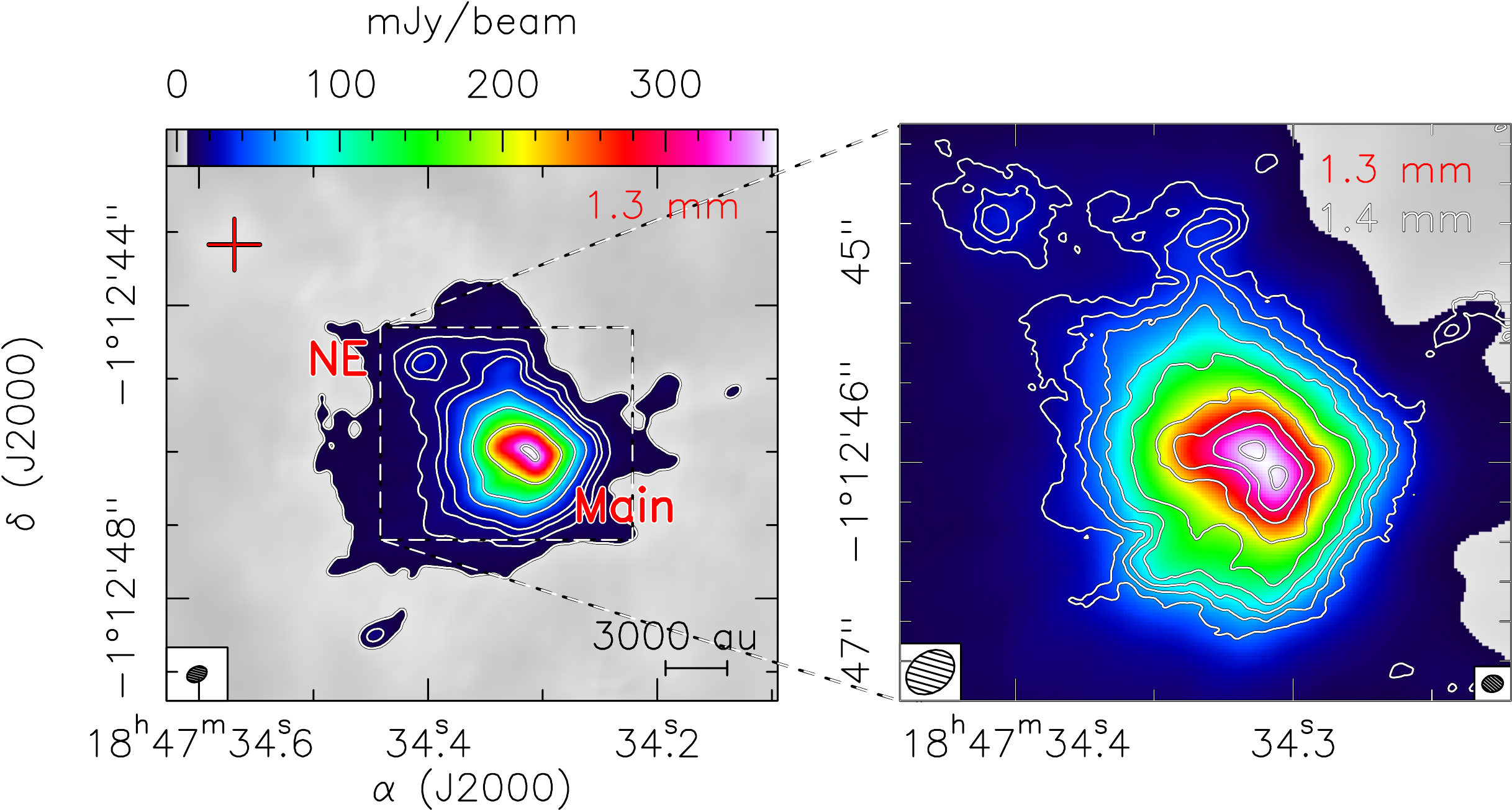}}
\caption{({\it Left}) ALMA map of the 1.3\,mm continuum emission of the G31 HMC from Beltr\'an et al.~(\cite{beltran19}). The contours are $-$5, 5, 10, 15, 20, 40, 80, 160, and 300 times $\sigma$, where 1$\sigma$ is 1.2\,mJy\,beam$^{-1}$. The labels indicate the two cores, {\it NE} and {\it Main}, resolved by B18. The red cross indicates the position of the \UC\ region imaged by Cesaroni et al.~(\cite{cesa94}).
The synthesized beam of $\sim$0$\farcs$24 is shown in the lower left-hand corner. ({\it Right}) Close-up of the central region around cores {\it NE} and {\it Main} that shows the ALMA 1.3\,mm continuum emission 
image overlaid with the ALMA 1.4\,mm continuum emission ({\it white
contours}) observed with an angular resolution of $\sim$0$\farcs$1. White contours are 3, 6, 9, 15, 30, 60, 90, 120, and 165 times 1\,$\sigma$, which is 0.32\mjy. The ALMA 1.3\,mm and 1.4\,mm synthesized beam are shown in the lower left-hand and right-hand corner, respectively.}
\label{fig-panels}
\end{figure*}

\section{Observations}
\subsection{ALMA}
\label{sect-ALMA}

Interferometric observations of G31 at 1.4\,mm in Band 6 and 3.5\,mm in Band 3 were carried out with ALMA in Cycle 4 and Cycle 6 as part of projects 2016.1.00223.S (P.I.: M.\ Beltr\'an) and 2018.1.00252.S (P.I.: M.\ Beltr\'an). Observations at 1.4\,mm were carried out in one of the more extended configurations, C40-7. The 1.4\,mm observations were combined with those of the ALMA Cycle 2 project 2013.1.00489.S (P.I.: R. Cesaroni), which were observed in the extended C34-7/(6) configuration and with the same digital correlator configuration. We refer to Cesaroni et al.~(\cite{cesa17}) and B18 for detailed information on the Cycle~2 observations. Observations at 3.5\,mm were performed in the very extended C40-9 and C43-9 configurations, and in the more compact C40-6 and C43-6 configurations (see Table~\ref{table_parobs} for more information). Observations at both wavelengths were initially carried out in Cycle 4 (project 2016.1.00223.S), between Sep.\ 2016 and Sep.\ 2017. After the project was completed, post-processing calibration and analysis of the 3.5\,mm data revealed severe problems on the bandpass shape of all spectral windows for both extended and compact configurations. This raised serious doubts about the calibration and overall quality of the  continuum map, in particular of the fluxes of the sources. Re-processing of the data by ALMA resulted in a quality assessment failure (QA2$\_$FAIL) and the scheduling blocks were put back in the observing queue of Cycle 5 but unfortunately were not scheduled so they could not be re-observed. After proposing again in Cycle\,6, the new observations at 3.5\,mm, in both extended and compact configurations, were finally carried out as part of the new project 2018.1.00252.S. In early 2020, after a thorough re-calibration of the 3.5\,mm Cycle 4 data\footnote{The Cycle 4 data were re-calibrated using an updated version of CASA (5.4.0-70) with the respective update of the ALMA pipeline that had improved features compared to the initial assessments made. Further manual changes were made in the reduction, especially the line-free selections for continuum reduction after the already improved {\it findcont} process of the pipeline. The combined reprocessing resulted in corrected baselines of the spectral windows and a confirmed flux scale with respect to the Cycle 6 data.} by one of the co-authors of this work, L.\ Maud, the spectral baselines were finally properly corrected and these data have been used in combination with those of  Cycle 6 to obtain the continuum emission map at 3.5\,mm.

The ALMA 1.4\,mm C40-7 observations used baselines in the range 21–-3637\,m, providing sensitivity to structures $<$1$\farcs$4. The Cycle 2 C34-7/(6) observations used baselines in the range 40--1500\,m.
The digital correlator was configured in thirteen spectral windows (SPWs), with SPW0 having a bandwidth of 1875\,MHz and 3840 channels while the rest used a 234.38\,MHz bandwidth and 960 channels. This translates into a velocity resolution of $\sim$1.3\,km\,s$^{-1}$ for SPW0 and $\sim$0.66--0.77\,km\,s$^{-1}$ for the other SPWs.  At 3.5\,mm, the source was observed in compact and extended configurations. Because the data were observed twice in different cycles, the baselines and maximum recoverable scales vary even for the same configurations, so we report the different values for {\it i)} the extended configuration: baselines 41--12145\,m (C40-9) and 83--16196\,m (C43-9), and maximum recoverable scales $<$1$\farcs$75 (C40-9) and $<$0$\farcs$97 (C43-9); and  {\it ii)} the compact configuration: baselines 15--3247\,m (C40-6) and 41--3637\,m (C43-6), and maximum recoverable scales $<5\farcs$7 (C40-6) and $<$4$\farcs$4 (C43-6).  The digital correlator at 3.5\,mm was configured in four SPWs of 1875\,MHz and 3840 channels each, which translates into a velocity resolution of $\sim$3.4$-$4.0\,km\,s$^{-1}$.

The phase reference center of the observations was set to the position $\alpha$(J2000) = 18$^{\rm h}$ 47$^{\rm m}$ 34$\fs$315, $\delta$(J2000) = $-$01$^\circ$ 12$'$ 45$\farcs$9. Flux and bandpass calibrations were achieved through observations of J1751+0939 and J1924$-$2914, while gain calibration through observations of J1851+0035. The data were calibrated and imaged using the {\sc CASA}\footnote{The {\sc CASA} package is available at \url{http://casa.nrao.edu/}} software package. Following ALMA Memo\,\#599\footnote{\url{https://library.nrao.edu/public/memos/alma/main/memo599.pdf}}, we conservatively assumed that the uncertainties on the absolute flux density calibration were $\sim$6\% at 3.5\,mm and $\sim$10\% at 1.4\,mm. Maps were created with the ROBUST parameter of Briggs (1995) set equal to 0. Further imaging and analysis were done with the {\sc GILDAS}\footnote{The {\sc GILDAS} package is available at \url{http://www.iram.fr/IRAMFR/GILDAS}} software package.  The continuum was determined from SPW0, which is the broadest spectral window at 1.4\,mm, centered at 218\,GHz, and the first spectral window at 3.5\,mm, centered at 85.3\,GHz. The continuum was subtracted from the line emission using the {\sc STATCONT}\footnote{\url{http://www.astro.uni-koeln.de/~sanchez/statcont}} algorithm (S\'anchez-Monge et al.~\cite{sanchez-monge18}). The resulting synthesized CLEANed beam and rms noise of the continuum maps are given in Table~\ref{table_parobs}.

\begin{figure*}
\centerline{\includegraphics[angle=0,width=17cm,angle=0]{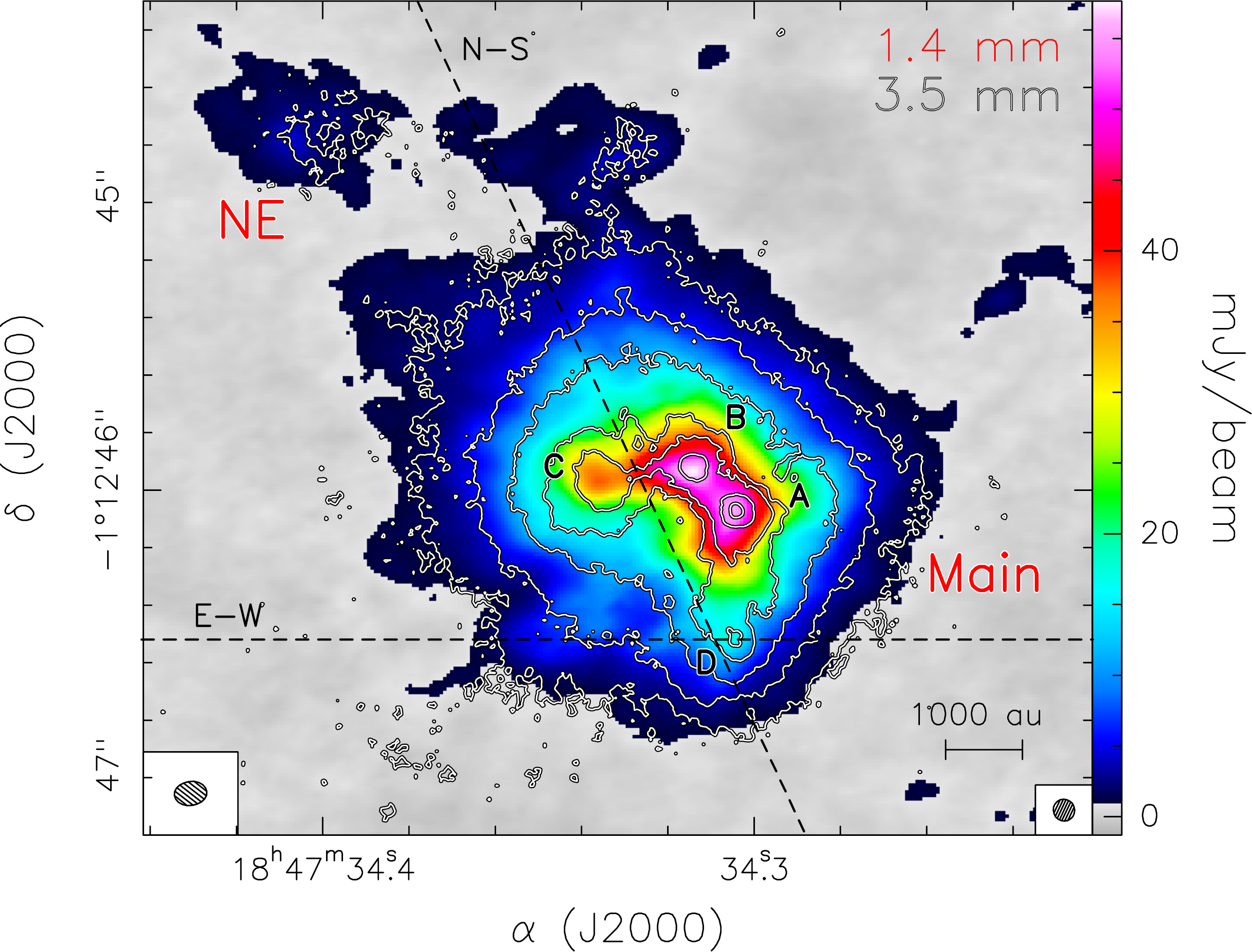}}
\caption{Overlay of the ALMA 3.5\,mm continuum emission ({\it white contours}) on the 1.4\,mm continuum emission ({\it colors}) toward cores {\it NE} and {\it Main} in G31. The white contours are 3, 6, 10, 15, 20, 30, and 40 times 1$\sigma$, which is 0.16\mjy. 
The 1.4\,mm and 3.5\,mm synthesized beams are shown in the lower left-hand and right-hand corner, respectively. The names of the sources in which the emission of the {\it Main} core has been resolved (see Sect.~\ref{sect-cont} and Table~\ref{table-cont}) are indicated. Dashed black lines indicate the direction of the E–W and N–S outflows mapped in SiO by B18.}
\label{fig-mm}
\end{figure*}

\subsection{VLA}
\label{sect-VLA}

Interferometric continuum observations of G31 at 1.3 cm (K-band) and 7 mm (Q-band) were carried out with the VLA of the National Radio Astronomy Observatory (NRAO) as part of the project 18A$-$031 (PI: V. M. Rivilla). The observations were done in the extended A-configuration between March and April 2018. The VLA K-band and Q-band observations used baselines in the range 680--36400\,m. We used the WIDAR backend with 2 GHz baseband pairs using the 3-bit samplers for both bands. The four basebands with 2.048\,GHz of bandwidth were centered at the frequencies 19.2, 21.2, 23.2, 25.2 GHz (K-band), and 41.0, 43.0, 45.0 and 47.0 GHz (Q-band). Each baseband contains 16 sub-bands of 128 MHz, with 64 channels of 2 MHz, which translates into a velocity resolution of 23--33\,km\,s$^{-1}$ in the K-band, and 12.5$-$15.0 km s$^{-1}$ in the Q-band. 
% in the range of frequencies covered
% Q-band: WIDAR backend correlator with 4 basebands of 2.048 GHz of bandwidth centered on 41.0, 43.0, 45.0 and 47.0 GHz (3-bit samplers). Each baseband contain 16 subbands of 128 MHz (64 channels of 2 MHz, which translates into 12.5-15.0 km s$^{-1}$ in the range of frequencies covered).
% 2 GHz baseband pairs when using the 3-bit samplers
% Full polarisation
% BI. BPs1	Recirculation 1
% Correlator Integration Time of 2 s.
% Observing time: 03:37 x 2
%%%%%%%
% Flux calibrator: 1331+305
% BP calibrator: 1733-1304 (NRAO530)
% Gain calibrator: 1851+0035
% Observing loop: 55 s on calibrator, 35 s on source, ; cycle times of 1.5 min.
%%%%%%%
%%%%%%%
%%%%%%%
%%%%%%%
% K-band: WIDAR backend correlator with 4 basebands of 2.048 GHz of bandwidth centered on 19.2, 21.2, 23.2, 25.2, and  GHz (3-bit samplers). Each baseband contain 16 subbands of 128 MHz (64 channels of 2 MHz, which translates into 23-33 km s$^{-1}$ in the range of frequencies covered).
% Observing time:03:13 x 2
% Full polarisation
% BI. BPs1	Recirculation 1
% Correlator Integration Time of 2 s.
%%%%%%%
% Flux calibrator: 1331+305
% BP calibrator: 1733-1304 (NRAO530)
% Gain calibrator: 1851+0035
% Observing loop:  55 s on calibrator, 2min 5s s on source; cycle times of 3 min
%
%
%The correlator was configured to cover X bands of X GHz, number of channels, spectral resolution.
The phase reference center of the observations was set to the position $\alpha$(J2000) = 18$^{\rm h}$ 47$^{\rm m}$ 34$\fs$315, $\delta$(J2000) = $-01^\circ$ 12$'$ 45$\farcs$90. Flux, bandpass, and gain calibrations were achieved respectively through observations of 1331+305, 1733-1304 (NRAO530), and  1851+0035. The gain-calibration cycles lasted 3 and 1.5 minutes for the K-band and Q-band observations, respectively.
The data were calibrated and imaged using the VLA Pipeline CASA software package\footnote{\url{https://science.nrao.edu/facilities/vla/data-processing/pipeline}}, which is designed to perform automatic flagging and calibration for Stokes I continuum data. Conservatively, the absolute accuracy of the flux density scale is estimated to be $\sim$5\% at both wavelengths (Perley \& Butler~\cite{perley17}). Further imaging and analysis was done with the {\sc GILDAS} package. The resulting synthesized beam and rms noise of the continuum maps are given in Table~\ref{table_parobs}.

By comparing the VLA and ALMA maps, we realized 
that there was a systematic positional offset between them. Since both observations used the same phase
calibrator 1851+0035 (J1851+0035), we cross-checked its coordinates in the two datasets and found an offset of $\Delta \alpha =$0$\farcs$021, $\Delta \delta =-$0$\farcs$0504. After contacting the VLA and ALMA staff, we reached the conclusion that the coordinates of the calibrator used by ALMA were more accurate than those used by the VLA. Therefore, the phase reference center of the VLA maps was shifted accordingly to the measured offset. After applying this correction, no significant offset was observed between the ALMA and VLA data.

%corrected the VLA maps taking as a reference the coordinates of the phase calibrator as reported by ALMA. In this way, the phase reference center of the VLA maps was shifted 0$\farcs$021 eastwards and $-$0$\farcs$0504 southwards. After applying this correction, no significant offset was observed between the ALMA and VLA data.
%Since the VLA position of 1851+0035 is only accurate within 0.15", according to the VLA Calibrator Manual\footnote{http://www.vla.nrao.edu/astro/calib/manual/csource.html},  we shifted the nominal position reported in the VLA catalog to match the ALMA coordinates. All the VLA maps  reported in this paper were shifted accordingly to the measured offset. 

%In fact, for all sources detected with both %instruments there was an offset in their positions that was always in the same direction. We checked the coordinates of the phase calibrator 1851+0035 (J1851+0035) used in the observations, which is the same for both VLA and ALMA, and found a discrepancy. The ALMA coordinates of the phase calibrator are $\alpha$(J2000)=18$^{\rm h}$ 51$^{\rm m}$ 46$\fs$7231,  
%$\delta$(J2000)=$+0^\circ$ 35$'$ 32$\farcs$3636, while those of VLA are $\alpha$(J2000)=18$^{\rm h}$ 51$^{\rm m}$ 46$\fs$7217, $\delta$(J2000)=$+0^\circ$ 35$'$ 32$\farcs$414.  

The VLA A-Array at K-band resolves out the more extended emission from the UC~\HII region located   $\sim$5$''$ away from the HMC (Cesaroni et al.~\cite{cesa94}), and, owing to sidelobe effects and poor cleaning, spurious extended features contaminate the K-band image of the HMC. First, we filtered out this spurious extended emission by cutting away all baselines with length < 5~km. Then, we tried to improve the dynamical range of the K-band map by self-calibrating the emission of the two compact sources inside the HMC. Phase-only self-calibration improved the dynamical range of the image only marginally (a few percent), while the signal is too weak for amplitude self-calibration. Thus, our final K-band image has no self-calibration applied. We also attempted to self-calibrate the compact emission at Q-band, but the signal-to-noise ratio is not sufficient even for phase-only self-calibration.

\begin{table*}
\caption[] {Position, flux densities, peak brightness temperature, and size of the sources embedded in the {\it Main} core of G31.41+0.31.}
\label{table-cont}
\begin{tabular}{lcccccccc}
\hline
&\multicolumn{2}{c}{Peak position$^a$}
&&&&\multicolumn{3}{c}{Source size$^c$}
\\
 \cline{2-3} 
 \cline{7-9} 
\multicolumn{1}{c}{$\lambda$} 
&\multicolumn{1}{c}{$\alpha({\rm J2000})$} &
\multicolumn{1}{c}{$\delta({\rm J2000})$} &
\multicolumn{1}{c}{$I^{{\rm peak}, b}_\nu$} &
\multicolumn{1}{c}{$T_{\rm B}$} &
\multicolumn{1}{c}{$S_\nu^c$} &
\multicolumn{1}{c}{FWHM$^{d}$} &
\multicolumn{1}{c}{P.\ A.} &
\multicolumn{1}{c}{$D_s^{e}$}  
\\
\multicolumn{1}{c}{(mm)} &
\multicolumn{1}{c}{h m s}&
\multicolumn{1}{c}{$\degr$ $\arcmin$ $\arcsec$} &
\multicolumn{1}{c}{(mJy/beam)} & 
\multicolumn{1}{c}{(K)} & 
\multicolumn{1}{c}{(mJy)} &
\multicolumn{1}{c}{(mas)} &
\multicolumn{1}{c}{(deg)} &
\multicolumn{1}{c}{(au)} 
\\
\hline
&&&& Source A \\
\hline
1.4  &18 47 34.305 &$-$01 12 46.07  &55.6$\pm$6   &150$\pm$15 &945$\pm$95  &447$\pm$2$\times$341$\pm$1 &161$\pm$1  &1445$\pm$7  \\
3.5  &18 47 34.304 &$-$01 12 46.07  &7.0$\pm$0.4   &206$\pm$12 &54$\pm$3  &236$\pm$5$\times$212$\pm$4 &60$\pm$8  &828$\pm$19  \\
7.0  &18 47 34.304 &$-$01 12 46.07  &0.79$\pm$0.04   &227$\pm$11 &2.0$\pm$0.10  &77$\pm$5$\times$56$\pm$4 &50$\pm$10  &243$\pm$19  \\
13.0  &18 47 34.304 &$-$01 12 46.07  &0.46$\pm$0.02   &225$\pm$11 &0.74$\pm$0.04  &84$\pm$2$\times$30$\pm$2 &6$\pm$1  &186$\pm$7  \\
\hline
&&&& Source B \\
\hline
%&\dittostraight
1.4  &18 47 34.314 &$-$01 12 45.94  &57.6$\pm$6   &156$\pm$16  &934$\pm$93  &465$\pm$2$\times$327$\pm$1 &83$\pm$1  &1443$\pm$7  \\
3.5     &18 47 34.314 &$-$01 12 45.92  &6.0$\pm$0.4   &178$\pm$11  &49$\pm$3 &224$\pm$6$\times$198$\pm$5 &37$\pm$9  &780$\pm22$  \\
7.0 &18 47 34.314 &$-$01 12 45.94  &0.56$\pm$0.03   &160$\pm$8  &1.2$\pm$0.1  &$61\pm$8$\times$55$\pm$9 &153$\pm$82  &215$\pm$33  \\
13.0  &18 47 34.314 &$-$01 12 45.94  &0.28$\pm$0.014   &139$\pm$7  &0.40$\pm$0.02  &56$\pm$4$\times$47$\pm$4 &170$\pm$25  &190$\pm$15  \\
\hline
&&&& Source C \\
\hline
1.4 &18 47 34.336 &$-$01 12 45.97  &37.6$\pm$4   &102$\pm$10  &464$\pm$46  &382$\pm$3$\times$292$\pm$2 &73$\pm$1  &1260$\pm$11  \\
3.5     &18 47 34.334 &$-$01 12 45.98  &4.8$\pm$0.3   &141$\pm$8  &67$\pm$4  &306$\pm$8$\times$273$\pm$7 &37$\pm$10  &1069$\pm$30  \\
7.0 &18 47 34.335 &$-$01 12 45.99  &0.14$\pm$0.03   &40$\pm$7  &0.6$\pm$0.14  &106$\pm$28$\times$83$\pm$24 &3$\pm$41  &347$\pm$104  \\
13.0  &--- &---  &$<$0.015   &$<$7  &$<$0.015  &--- &--- &---  \\
\hline
&&&& Source D \\
\hline
1.4 &18 47 34.306 &$-$01 12 46.51  &17.7$\pm$2   &48$\pm$5  &166$\pm$17  &323$\pm$5$\times$243$\pm$4 &173$\pm$3  &1036$\pm$19  \\
3.5     &18 47 34.304 &$-$01 12 46.51  &3.5$\pm$0.2   &102$\pm$6  &38$\pm$2  &287$\pm$11$\times$212$\pm$9 &22$\pm$5  &912$\pm$41  \\
7.0 &18 47 34.305 &$-$01 12 46.52  &0.16$\pm$0.03   &47$\pm$7  &0.7$\pm$0.14  &127$\pm$27$\times$68$\pm$18 &104$\pm$14  &344$\pm$100  \\
13.0  &18 47 34.304 &$-$01 12 46.53  &0.04$\pm$0.010   &19$\pm$2  &0.04$\pm$0.010  &<71$^f$ &---  &$<263^f$ \\
\hline
\end{tabular}
\tiny

  $^a$ Position of the dust emission peak at each wavelength estimated from the maps.\\
  $^b$ Peak intensity and integrated flux density corrected for primary beam
  response. For the peak intensity, we adopted the most conservative error, either the uncertainty on the flux calibration or the rms noise of the map. For the integrated flux density, we adopted the most conservative error, either the uncertainty on the flux calibration or the error of the 2-D Gaussian fit.   \\
  $^c$ Estimated by fitting elliptical 2-D Gaussians. \\
  $^d$ Deconvolved Full Width Half Maximum (FWHM). \\
  $^e$ Deconvolved geometric mean diameter.  The distance to G31 is assumed to be 3.75\,kpc (Immer et al.~\cite{immer19}).\\ 
  $f$ Source unresolved.  
\end{table*}

\begin{figure*}
%\hspace{.5cm}
\centerline{\includegraphics[angle=0,width=17cm,angle=0]{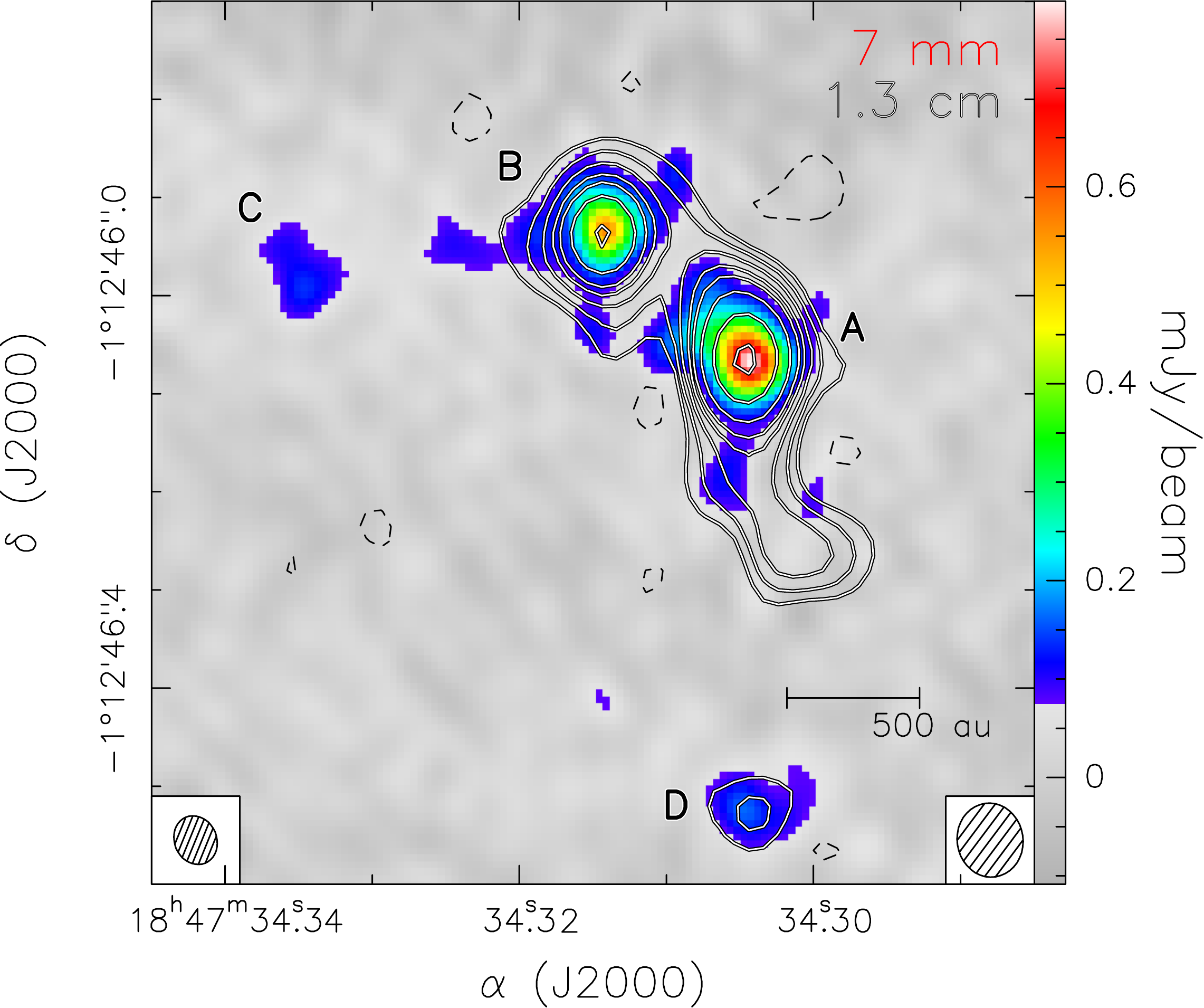}}
%\centerline{\includegraphics[angle=0,width=17cm,angle=0]{masers-7mm-1cm.eps}}
\caption{Overlay of the VLA 1.3\,cm continuum emission ({\it contours}) on the 7\,mm continuum emission ({\it colors}) toward cores {\it NE} and {\it Main} in G31. The  dashed black contours are $-$3 and the white contours 3, 6, 10, 15, 20, 30, 50, and 80 times 1$\sigma$, which is  0.005\mjy.  The 1.3\,cm and 7\,mm synthesized beams are shown in the lower left-hand and right-hand corner, respectively. The names of the cores in which the emission of the {\it Main} core has been resolved are indicated.}
\label{fig-cm}
\end{figure*}

\begin{figure*}
%\hspace{.5cm}
\centerline{\includegraphics[angle=0,width=17cm,angle=0]{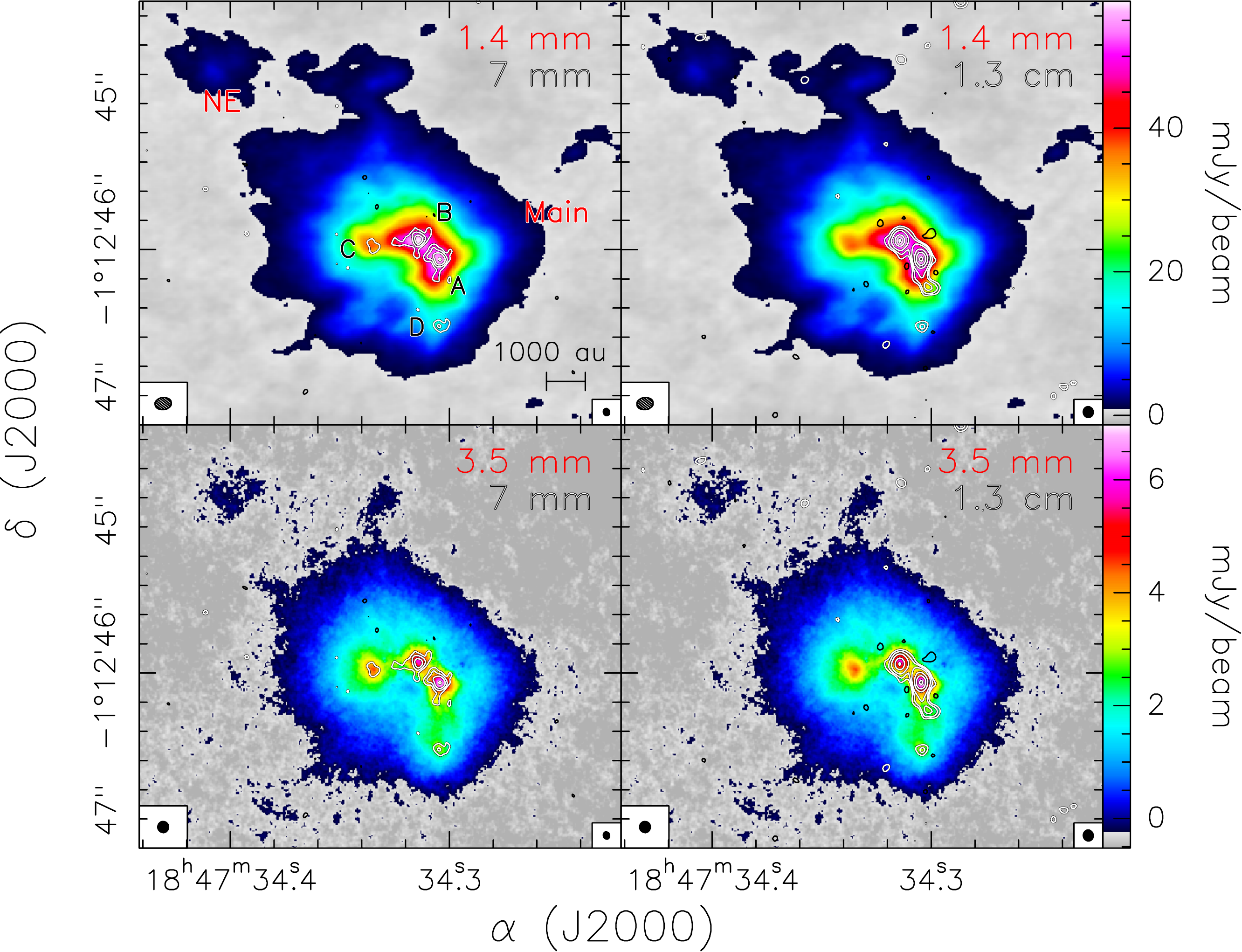}}
\caption{({\it Top panels}) Overlay of the VLA 7\,mm ({\it left}) and 1.3\,cm ({\it right})
continuum emission ({\it contours})  on the ALMA 1.4\,mm continuum emission ({\it colors}) toward cores {\it NE} and {\it Main} core in G31. The  black contours are $-$3 and the white contours 3, 6, 12, 30, 50, and 80 times 1$\sigma$, which is 0.025\mjy\ at 7\,mm and 0.005\mjy\ at 1.3\,cm. The 1.4\,mm synthesized beam is shown in the lower left-hand corner of the panels, and the 7\,mm and 1.3\,cm synthesized beams in the lower right-hand corner. The names of the sources in which the emission of the {\it Main} core has been resolved are indicated. ({\it Bottom panels}) Same but overlaid on the ALMA 3.5\,mm continuum emission ({\it colors}).}
\label{fig-mm-cm}
\end{figure*}

\section{Results and Analysis}

\subsection{Continuum emission of the {\it Main} and {\it NE} cores}
\label{sect-cont}

Figure~\ref{fig-panels} ({\it left panel}) shows the dust continuum emission of the G31 HMC at 1.3\,mm obtained by Beltr\'an et al.~(\cite{beltran19}) in their dust polarized emission observations carried out with ALMA at an angular resolution of $\sim$0$\farcs$24, which have clearly resolved the emission of G31 into two cores, {\it Main} and {\it NE}. In this figure, the {\it Main} core shows a monolithic and homogeneous appearance.  The right panel shows the new high angular resolution ($\sim$0$\farcs1$) observations at 1.4\,mm overlaid on the  $\sim$0$\farcs$24 observations. In the higher angular resolution map, the emission from both cores is clearly detected and the emission of the {\it Main} core starts to show a less homogeneous morphology revealing the presence of clumpiness. In addition, the observations have also revealed an elongation to the north of the {\it Main} core that was already visible but not previously resolved. 

The structure of the {\it Main} core is better highlighted at 3.5\,mm as shown in Fig.~\ref{fig-mm}. At this wavelength and with slightly higher angular resolution ($\sim$0$\farcs075$), the emission is clearly resolved into four embedded cores, labelled A, B (following the nomenclature of C10), C, and D. To avoid confusion between the larger cores observed with an angular resolution of $\sim$0$\farcs24$ and the cores resolved with higher angular resolution, hereafter we will use the term ``source'' for the latter. Sources A and B, which are located at the center of the core and are separated by $\sim$0$\farcs$20 ($\sim$750\,au), coincide with the two free-free continuum emission sources first detected at 1.3\,cm and 7\,mm by C10. The morphologies of these two sources are quite similar to each other. Source C lies $\sim$0$\farcs$30 ($\sim$1100\,au) east of source B, while source D is located $\sim$0$\farcs$45 ($\sim$1700\,au) south of source A. This latter source is located close to the intersection of two of the SiO outflows, the N--S and the E--W, detected by B18 (see Fig.~\ref{fig-mm}). The fact that the source is located along the direction of the E--W SiO outflow and close to its center of symmetry suggests that source D is likely its driving source. The {\it NE} core and the northern elongation are marginally visible at 3.5\,mm. 

Figure~\ref{fig-cm} shows the continuum emission at 1.3\,cm and 7\,mm obtained with new VLA observations, which besides detecting sources A and B observed by C10, have also detected the emission from sources C and D at 7\,mm, while at 1.3\,cm only source D has been detected. At 1.3\,cm, the emission of the sources is quite compact, with the exception of source A that shows a central compact component plus an elongation to the south. This elongation was marginally visible with the 1.3\,cm observations of C10. Figure~\ref{fig-mm-cm} shows the VLA 7\,mm and 1.3\,cm continuum emission overlaid on the ALMA 1.4\,mm and 3.5\,mm continuum emission. The agreement in the position of the sources at the different wavelengths is remarkable. We note that neither the {\it NE} core nor the elongation observed northward of the {\it Main} core at 1.4\,mm and 3.5\,mm have been detected at 7\,mm or 1.3\,cm.

\subsection{Parameters of the sources embedded in the {\it Main} core}
\label{sect-param}

\begin{figure*}
%\hspace{.5cm}
\centerline{\includegraphics[angle=0,width=17cm,angle=0]{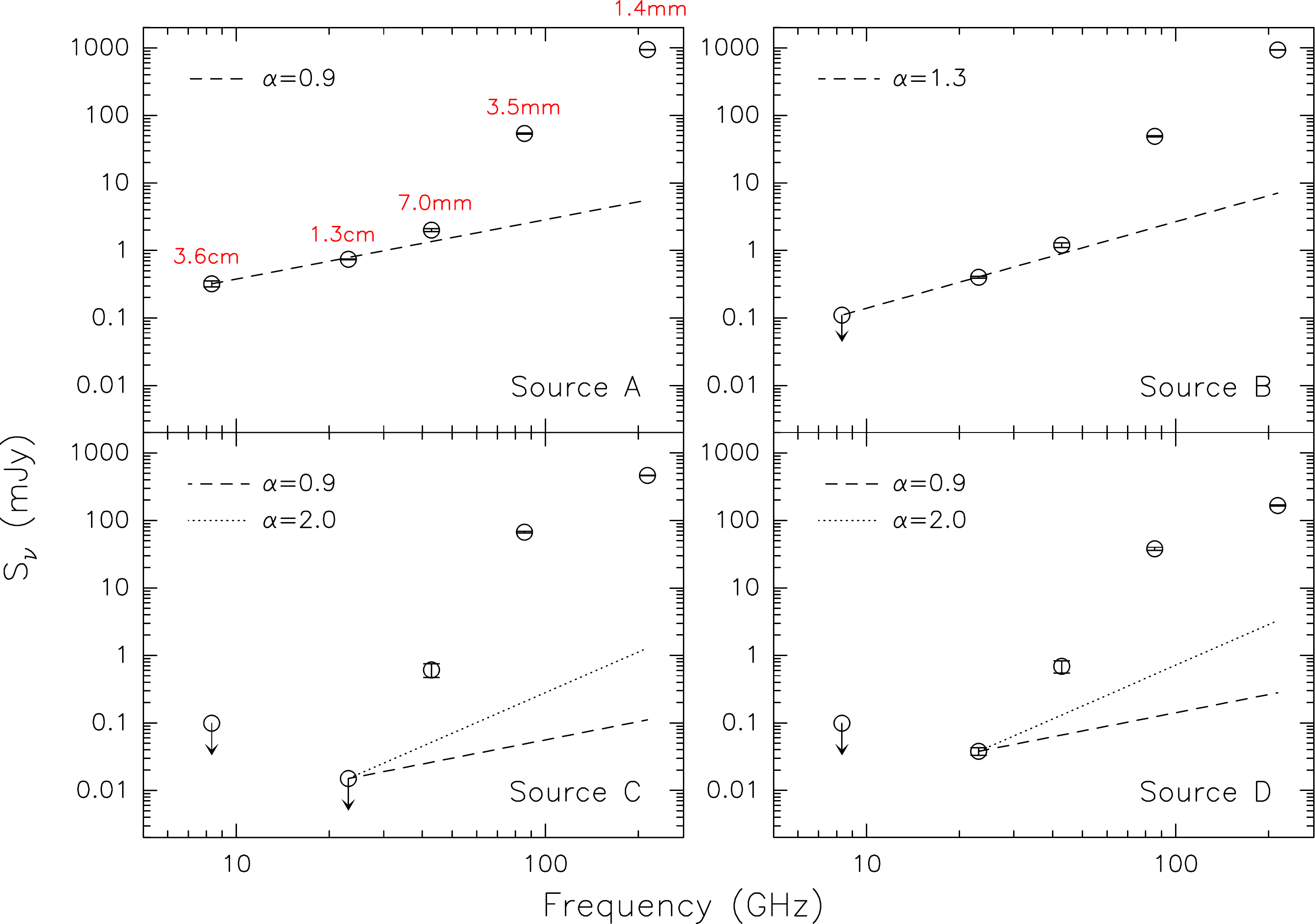}}
\caption{Spectral energy distribution of the four continuum sources detected in the {\it Main} core of G31. The dashed lines indicate the extrapolation of the fluxes at 3.6\,cm or 1.3\,cm with a power law of the type $S_\nu\propto \nu^\alpha$. The values of $\alpha$ are given in each box.}
\label{fig-sed}
\end{figure*}

Table~\ref{table-cont} lists the position, flux density, and size of the four sources embedded in the {\it Main} core of G31 at the four different wavelengths. The positions of the emission peaks of each source at the different wavelengths coincide. Furthermore, sources A and B are the brightest at all wavelengths, and have similar peak intensities at millimeter wavelengths, while at 1.3\,cm, source A is twice as bright as source B. The integrated flux density of the sources at the different wavelengths has been estimated by fitting elliptical 2-D Gaussians in the image plane. 
%, we have taken into account how well resolved their emission was from the other sources and from the more extended core envelope. At 1.3\,cm and 7\,mm, the emission of the sources is properly resolved, and therefore, their integrated flux has been estimated inside a 3$\sigma$ level (see Table~\ref{table-cont}). At 3.5\,mm and especially at 1.4\,mm, the emission of the different sources is more difficult to resolve at low emission levels (e.g., 3$\sigma$), and it is also difficult to disentangle the emission of the sources from that of the common envelope that enshroud all them. Taking a look at the 3.5\,mm map, one sees that the best contour level to separate the emission of the four different sources is that at 15$\sigma$, and this is the contour that we have used to estimate their fluxes (see Table~\ref{table-cont}). At 1.4\,mm, the emission of the sources is better resolved at a contour level of 60$\sigma$, which means that the fluxes estimated in Table~\ref{table-cont} should be probably considered a lower limit. For the weaker source D, the contour level used to calculate the integrated flux is 30$\sigma$. 
The spectral energy distributions (SEDs) of the sources are shown in Fig.~\ref{fig-sed}. At 1.4\,mm, 7\,mm, and 1.3\,cm, the integrated flux densities of sources A and B are clearly greater than those of sources C and D, but at 3.5\,mm, the emission of source C is stronger than that of sources A and B. 
%This is probably due to the fact that the size of source C at 3\,mm is larger than that of sources A and B.

The size of the sources has been estimated by fitting elliptical 2-D Gaussians in the image plane. At the highest angular resolution achieved, that is $\sim$0$\farcs$05 (or $\sim$180\,au at the distance of G31) at 7\,mm, all the sources have been resolved and have deconvolved geometric mean diameters of $\sim$200--350\,au. At any wavelength but 1.3\,cm, the sizes of sources A and B are comparable. At 1.3\,cm, the major axis of source A is greater than that of source B,  probably due to the presence of the southern elongation visible in the maps (see Fig.~\ref{fig-cm}) which could be associated with source A or could indicate the presence of an additional source. Therefore, the similarities between sources A and B are not only related to their peak intensities and integrated flux densities but also to their morphologies and sizes. 

\subsubsection{Free-free contribution to the emission} 

\label{sect-free}

As already mentioned, C10 observed the region at 3.6\,cm, 1.3\,cm, and 7\,mm and resolved the emission into two sources named A and B. These authors discussed the nature of their  continuum emission in terms of \HII regions or thermal radio jets, and discarded the former hypothesis. An important outcome of the new data is the good match between the positions of the centimeter and the millimeter peaks revealed by the new VLA and ALMA observations (Fig.~\ref{fig-mm-cm}). Such a good match proves that the free-free emission from A and B is tracing two distinct (jets from) young stellar objects (YSOs) and not the lobes of the same radio jet. In fact, in the latter case one should see only one millimeter peak located right in the middle of the two free-free peaks. 

\begin{table*}
\caption[] {Free-free continuum flux densities$^a$.} 
\label{table-cm}
\begin{tabular}{lccccc}
\hline
&\multicolumn{1}{c}{$S_{\rm 1.4mm}$}
&\multicolumn{1}{c}{$S_{\rm 3.5mm}$}
&\multicolumn{1}{c}{$S_{\rm 7.0mm}$}
&\multicolumn{1}{c}{$S_{\rm 1.3cm}$} 
&\multicolumn{1}{c}{$S_{\rm 3.6cm}$} 
\\
\multicolumn{1}{c}{Source} 
&\multicolumn{1}{c}{(mJy)} 
&\multicolumn{1}{c}{(mJy)} 
&\multicolumn{1}{c}{(mJy)} 
&\multicolumn{1}{c}{(mJy)} 
&\multicolumn{1}{c}{(mJy)} 
\\
\hline
A &5.8(0.6\%)$^b$ &2.5(5\%)$^b$ &1.4(70\%)$^b$ &0.74(100\%) &0.32(100\%) \\
B &7.8(0.8\%)$^c$ &2.3(5\%)$^c$ &1.0(83\%)$^c$ &0.40(100\%) &$<0.11$(100\%) \\
C &0.12(0.03\%)--1.4(0.3\%)$^d$ &0.05(0.1\%)--0.22(0.3\%)$^d$ &0.03(5\%)--0.06(10\%)$^d$ &$<0.015$(100\%) &$<0.099$(100\%) \\
D &0.3(0.2\%)--3.7(2\%)$^d$ &0.13(0.3\%)--0.56(1.4\%)$^d$ &0.07(10\%)--0.15(21\%)$^d$ &0.038(100\%)  &$<0.099$(100\%)\\
\hline
\end{tabular}
\tiny

$^a$ The percentages in parenthesis indicate the contribution of the free-free emission to the total emission of the source. \\  
$^b$ Estimated by extrapolating the flux at 1.3\,cm with a spectral index of 0.9 (see Sect.~\ref{sect-free}). \\
$^c$ Estimated by extrapolating the flux at 1.3\,cm with a spectral index of 1.3 (see Sect.~\ref{sect-free}). \\
$^d$ Range of flux densities estimated by extrapolating the flux at 1.3\,cm with a spectral index of 0.9 and 2.0 (see Sect.~\ref{sect-free}). \\
\end{table*}

C10 fitted the SED of sources A and B assuming that the emission originates from thermal radio jets with a power law $S_\nu\propto\nu^\alpha$, where the spectral index $\alpha$ is 0.9 for source A and 1.3 for source B. These spectral indices are consistent with the values found for partially thick  thermal radio jets (e.g., Anglada et al.~\cite{anglada18}). The integrated flux emission, $S_\nu$, at 1.3\,cm estimated by C10 for sources A and B is 0.76 and 0.46\,mJy, respectively, and is consistent with the values measured by us (see Table~\ref{table-cont}). Therefore, using the same spectral index of 0.9 for source A and 1.3 for source B and extrapolating the 1.3\,cm fluxes to 7\,mm, 3.5\,mm, and 1.4\,mm, we estimated the expected free-free emission of the sources. We used the 1.3\,cm flux instead of the 3.6\,cm because the emission of the sources is better resolved and the angular resolution at 1.3\,cm, with a synthesized beam of 0$\farcs$095$\times$0$\farcs$084 (C10) is comparable to that of the ALMA observations (0$\farcs$075--0$\farcs$1). Table~\ref{table-cm} shows the expected flux densities at the different wavelengths as well as the contribution of the free-free emission to the total emission. 
The contribution of the free-free emission for sources A and B is important at 7\,mm and negligible at 3.5\,mm and 1.4\,mm. Figure~\ref{fig-sed} shows the SEDs of the sources, with a dashed line indicating the extrapolation of the  1.3\,cm fluxes with a power law with an index of 0.9 for source A and of 1.3 for source B. 

The sensitivity of our VLA observations is $>6$ times higher at 1.3\,cm and $>3$ times at 7\,mm than that of C10, and this allowed us to detect source D at both wavelengths, and source C at 7\,mm. With the current data, it is not possible to estimate the spectral index of the free-free emission for source C. For source D, despite the fact that the source has been detected at both 1.3\,cm and 7\,mm, an accurate estimate of the spectral index is not possible because the emission at 7\,mm could be contaminated by thermal dust emission. We therefore estimated the expected free-free flux by assuming two different spectral indices: $\alpha=2$, for which the contribution of the free-free emission at millimeter wavelengths should be maximum; and $\alpha$=0.9, the lowest spectral index estimated by C10 for the sources embedded in the {\it Main} core, which should give the lowest free-free contribution at millimeter wavelengths. Note that if the spectral index were $<$0.9, the free-free contribution at millimeter wavelengths would be even smaller. Table~\ref{table-cm} gives the expected free-free fluxes and the contribution to the total flux emission, while in Fig.~\ref{fig-sed}, we show with a dotted and a dashed lines the extrapolations of the 1.3\,cm emission for $\alpha=2$ and 0.9, respectively. Taking into account that source D is probably the driving source of the SiO E--W outflow detected by B18, then its free-free emission is likely due to a thermal radio jet, which has $\alpha<1.3$ (see Anglada et al.~\cite{anglada18} and references therein). In any case, independently of the spectral index used to extrapolate the centimeter emission to millimeter wavelengths, the expected free-free contribution at 1.4\,mm and 3.5\,mm is negligible.

In conclusion, while the continuum emission at 7\,mm still has an important contribution of free-free emission, especially for sources A and B, the emission at 3.5\,mm and 1.4\,mm is dominated by thermal dust emission for all sources.

\begin{table*}
\caption[] {Parameters of the sources embedded in the {\it Main} core$^a$.}
\label{table-mass}
\begin{tabular}{lccccccccc}
\hline
&\multicolumn{1}{c}{{\it offset}$^b$}
&\multicolumn{1}{c}{$T_{\rm ex}^c$}
&\multicolumn{1}{c}{$R^d$}
&\multicolumn{1}{c}{$M_{\rm gas}^e$} 
&\multicolumn{1}{c}{$N_{\rm H_2}^f$} 
&\multicolumn{1}{c}{$n_{\rm H_2}^f$} 
&\multicolumn{1}{c}{$\Sigma^g$} 
\\
\multicolumn{1}{c}{Source} 
&\multicolumn{1}{c}{(arcsec)} 
&\multicolumn{1}{c}{(K)} 
&\multicolumn{1}{c}{(au)}
&\multicolumn{1}{c}{($M_\odot$)} 
&\multicolumn{1}{c}{(10$^{25}$\,cm$^{-2}$)} 
&\multicolumn{1}{c}{(10$^{9}$\,cm$^{-3}$)} 
&\multicolumn{1}{c}{(g\,cm$^{-2}$)}
&\multicolumn{1}{c}{$\tau_{\rm 3.5mm}^h$}
&\multicolumn{1}{c}{$\tau_{\rm 1.4mm}^h$}
\\
\hline
A &0.22  &500 &414  &16$\pm$3  &5.8$\pm$1.0 &7.0$\pm$1.2 &270$\pm$45 &0.5  &2.7 \\
B &0.04  &500 &390  &15$\pm$3  &5.9$\pm$1.0 &7.6$\pm$1.3 &276$\pm$46 &0.6  &2.8 \\
C &0.30  &390 &535  &26$\pm$5  &5.5$\pm$1.1 &5.2$\pm$1.1 &257$\pm$53 &0.5  &2.6  \\
D &0.62  &225 &456  &26$\pm$8  &7.5$\pm$2.3 &8.2$\pm$2.5 &349$\pm$108 &0.7  &3.5\\
\hline
\end{tabular}
\tiny

$^a$ Estimated from the 3.5\,mm emission. \\
$^b$ Distance to the phase reference center of the observations of B18, $\alpha$(J2000) = 18$^{\rm h}$ 47$^{\rm m}$ 34$\fs$315, $\delta$(J2000) = $-01^\circ$ 12$'$ 45$\farcs$90, which is the central position of the temperature profile (see Sect.~\ref{section-mass}). \\
$^c$ Methyl formate excitation temperature, estimated following Eq.~(1) of B18 (see Sect.~\ref{section-mass}). \\
$^d$ Radius of the source obtained by fitting elliptical 2-D Gaussians (Table~\ref{table-cont}).\\
$^e$ Mass estimated using a dust opacity of 0.2\,cm$^2$\,g$^{-1}$ at 3.5\,mm and a dust temperature equal to the CH$_3$OCHO excitation temperature (see Sect.~\ref{section-mass}). \\
$^f$ Column density and number density estimated from $M_{\rm gas}$, assuming $\mu_{\rm H}$=2.8.\\
$^g$ Source mass surface density derived as $\Sigma = M_{\rm gas}/\pi\,R^2$, where $R$ is the radius of the source. \\
$^h$ Dust opacity depth estimated as $\tau=\mu_{\rm H}\,m_{\rm H}\,\kappa_\nu\,N_{\rm H_2}/100$, where $\kappa_\nu$ is the dust opacity, and 100 is the gas-to-dust mass ratio.
\end{table*}

\subsection{Mass estimates}
\label{section-mass}

The mass of the sources embedded in the {\it Main} core was estimated assuming that the emission is optically thin, by using
\begin{equation}
    M_{\rm gas} = \frac{g\,S_\nu\,d^2}{\kappa_\nu\,B_\nu(T_{\rm d})}
\end{equation}
where $S_\nu$ is the flux density, $d$ the distance to the source, $\kappa_\nu$ the dust opacity coefficient, $g$ the gas-to-dust ratio, and $B_\nu(T_{\rm d})$ the Planck function for a blackbody with dust temperature $T_{\rm d}$. 
We used the continuum emission at 3.5\,mm, because at this wavelength the emission is optically thinner than at 1.4\,mm. Based on the findings of Sect.~\ref{sect-free}, we assumed that the continuum emission at 3.5\,mm is pure dust thermal emission and estimated the masses, $M_{\rm gas}$, the mean H$_2$ column densities, $N_{\rm H_2}$, the mean H$_2$ volume densities, $n_{\rm H_2}$, and the mass surface density, $\Sigma$, of the four sources associated with the {\it Main} core (see
Table~\ref{table-mass}). We adopted a
dust opacity of 0.2\,cm$^2$\,g$^{-1}$ at 3.5\,mm (by extrapolating, with a dust opacity index $\beta$=1.8, the value at 1.3\,mm of Ossenkopf \& Henning~\cite{ossenkopf94} for a Mathis,  Rumpl, \& Nordsieck~(\cite{mathis77}; MRN) distribution with thin ice mantles and a gas density of 10$^8$\,cm$^{-3}$) and a gas-to-dust mass ratio of 100. Recent works propose higher values of the gas-to-dust mass ratio, e.g.\ 150 (Draine~\cite{draine11}) or 162 (Peters et al.~\cite{peters17}). One can estimate the masses and densities given in Table~\ref{table-mass} with a different gas-to-dust mass ratio simply by multiplying the values by (new gas-to-dust mass ratio)/100. Assuming that gas and dust are coupled, we used the excitation temperature, $T_{\rm ex}$, estimated for methyl formate, CH$_3$OCHO $v=0$ and $v_t=1$ by B18 as $T_{\rm d}$. These authors fitted the CH$_3$OCHO emission as a function of radius across the {\it Main} core thus obtaining a temperature profile. Therefore, from the distance of the source to the center of the {\it Main} core, it is possible to estimate $T_{\rm ex}$ using their Eq.~(1). The temperature profile of B18 is valid for radii $\geq$0$\farcs$22 (the angular resolution of their ALMA observations), therefore  we assumed that sources A and B, located at 0$\farcs22$ and 0$\farcs04$ from the central position of the temperature profile, have 
the same $T_{\rm ex}$ of $\sim$500\,K, estimated at  0$\farcs$22. Because the temperature profile was obtained with an angular resolution $\sim$3 times lower than that of the 3.5\,mm observations, part of the gas inside the beam 
could be slightly colder and $T_{\rm ex}$ is to be taken as a lower limit. This would decrease the mass and density estimates. In order to have an idea of the uncertainty introduced by the previous assumption on $T_{\rm ex}$, we have recomputed the masses and densities increasing the temperatures by 100\,K.   

The masses of the sources range from $\sim$15 to $\sim$26\,$M_\odot$ (Table~\ref{table-mass}), suggesting that the cores are probably massive enough to form high-mass young stellar objects. The H$_2$ column densities are higher than $10^{25}$\,cm$^{-2}$ for all sources. Such values correspond to visual extinctions of $>5000$\,mag, confirming that the sources are deeply embedded in the core. The volume densities are also very high, several 10$^9$\,cm$^{-3}$. Using the $N_{\rm H_2}$ values, we estimated the dust optical depth at 3.5\,mm, and obtained values between 0.5 and 0.7 (see Table~\ref{table-mass}). These opacity values indicate that at 3.5\,mm the dust emission is mostly optically thin and that the masses should not be largely underestimated.

\begin{figure*}
%\hspace{.5cm}
\centerline{\includegraphics[angle=0,width=17cm,angle=0]{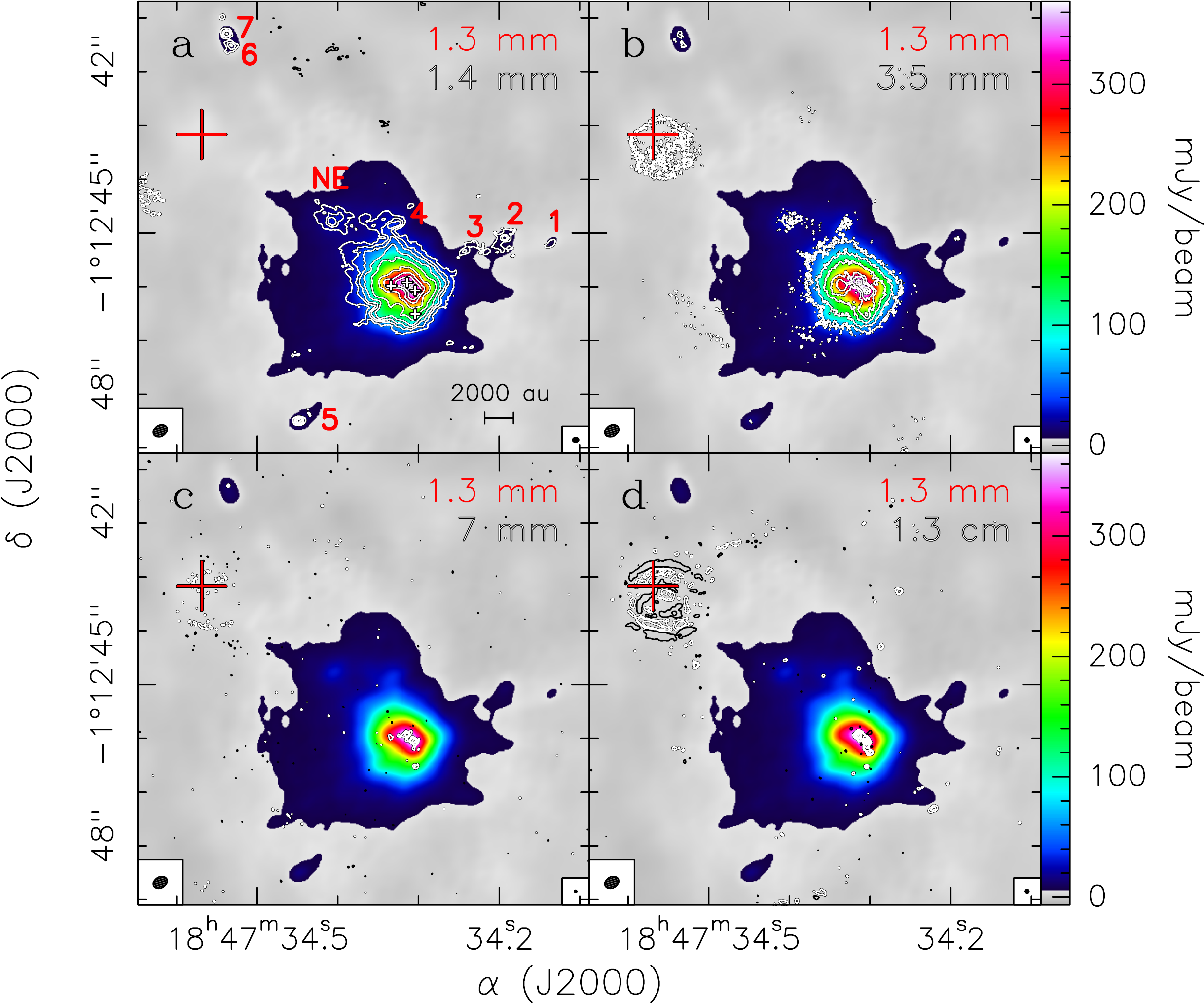}}
\caption{Overlay of the ALMA 1.4\,mm ({\it a}) and 3.5\,mm ({\it b}), and VLA 7\,mm ({\it c}) and 1.3\,cm ({\it d}) continuum emission ({\it contours}) on the ALMA 1.3\,mm continuum emission of the G31 HMC observed by Beltr\'an et al.~(\cite{beltran19}). White contour levels are the same as in Figs.~\ref{fig-panels}, \ref{fig-mm}, and \ref{fig-mm-cm}. Black contours are $-$3 times 1$\sigma$. The 1.3\,mm synthesized beam is shown in the lower left-hand corner and that at the other wavelengths in lower right-hand corner of each panel. The red cross marks the position of the \UC\ region imaged by Cesaroni et al.~(\cite{cesa94}),  while the white crosses indicate the positions of the four sources embedded in the {\it Main} core. The labels of the other sources detected in the region are indicated in red (see Sect.~\ref{sect-allcore}) and their positions given in Table~\ref{table-other}.}
\label{fig-allcore}
\end{figure*}

\subsection{Other sources in the region}
\label{sect-allcore}

Figure~\ref{fig-allcore} shows the $8\farcs5$ central region of the G31 HMC at the different wavelengths. The UC \HII region located $\sim$5$''$ northeast of the {\it Main} core is clearly detected at 1.3\,cm, 7\,mm, and 3.5\,mm, albeit partially resolved out. The UC \HII region appears spherical, with a diameter of $\sim$1$''$ or 3750\,au (0.018\,pc). According to C10, the star ionizing the UC \HII region would be an O6 star.

Besides the UC \HII region and the sources associated with the {\it Main} core and the {\it NE} core, our observations at millimeter wavelengths have revealed the presence of other sources in the region. The sources have been identified by eye. We consider a source real when detected at least at a 3$\sigma$ level and at two different wavelengths. This last constrain has the purpose of avoiding the spurious detection of weak sources which are actually noise. The mass sensitivity estimated at 3.5\,mm for a 3$\sigma$ detection limit is 1.5--4.0\,$M_\odot$, for a range of temperature of 20--50\,K. At 1.4\,mm, the 3$\sigma$ mass sensitivity is 0.1--0.3\,$M_\odot$ for the same temperature range.

Following what done for source A at 1.3\,cm, we have decided not to split the sources when elongated (e.g., source $NE$ or \#4), unless their emission clearly shows well separated peaks. The possible multiplicity of these sources will be better investigated with future higher-angular resolution observations. The positions of these other sources, the wavelengths at which they have been detected, and their flux densities at 1.4\,mm and 3.5\,mm can be found in Table~\ref{table-other}. The {\it NE} core has also been included in the table. We note that none of these are detected at centimeter wavelengths. The total number of sources detected in G31 is twelve plus the UC \HII region, but taking into account that some sources could be actually multiple, this number should be taken as a lower limit.

Table~\ref{table-other} shows the gas masses of these 7 additionally detected sources. The masses have been estimated from the continuum at 3.5\,mm except for two sources not detected at this wavelength for which the mass has been estimated at 1.4\,mm. We adopted a dust opacity of 1.0\,cm$^2$\,g$^{-1}$ at 1.4\,mm (Ossenkopf \& Henning~\cite{ossenkopf94} for a MRN distribution with thin ice mantles and a gas density of 10$^8$\,cm$^{-3}$).  For these sources, there are no temperature estimates. Therefore, to estimate the masses we have used a range of temperatures from 50 to 100\,K. The values of the masses (Table~\ref{table-other}) suggest that all but source \#4 and the {\it NE} core may form young stellar objects of low to intermediate mass.

\begin{table*}
\caption[] {Positions, flux densities and masses of other sources in the region}
\label{table-other}
\begin{tabular}{lcccccc}
\hline
&\multicolumn{2}{c}{Position}
\\
 \cline{2-3} 
&\multicolumn{1}{c}{$\alpha({\rm J2000})$} &
\multicolumn{1}{c}{$\delta({\rm J2000})$} &
\multicolumn{1}{c}{$\lambda^a$}  &
\multicolumn{1}{c}{$S_{\rm 1.4mm}$} &
\multicolumn{1}{c}{$S_{\rm 3.5mm}$} &
\multicolumn{1}{c}{$M_{\rm gas}^b$} 
\\
\multicolumn{1}{c}{Source} &
\multicolumn{1}{c}{h m s}&
\multicolumn{1}{c}{$\degr$ $\arcmin$ $\arcsec$} &
\multicolumn{1}{c}{(mm)} &
\multicolumn{1}{c}{(mJy)} & 
\multicolumn{1}{c}{(mJy)} & 
\multicolumn{1}{c}{($M_\odot$)} 
\\
\hline
1 &18 47 34.137 &$-$01 12 45.20 &1.3, 1.4 &2.5$\pm$0.3 &$<0.5$ &0.12--0.3$^c$\\
2 &18 47 34.191 &$-$01 12 45.10 &1.3, 1.4, 3.5 &11$\pm$0.3 &0.7$\pm$0.2 &1--2\\
3 &18 47 34.243 &$-$01 12 45.34 &1.3, 1.4 &3.0$\pm$0.3 &$<0.5$ &0.2--0.3$^c$\\
4 &18 47 34.334 &$-$01 12 44.86 &1.3, 1.4, 3.5 &30$\pm$0.3 &5.3$\pm$0.2 &8--17\\
5 &18 47 34.449 &$-$01 12 48.49 &1.3, 1.4, 3.5 &7.9$\pm$0.3 &0.5$\pm$0.2 &0.8--2\\
6 &18 47 34.533 &$-$01 12 41.51 &1.3, 1.4, 3.5  &7.7$\pm$0.3 &1.2$\pm$0.2 &2--4\\
7 &18 47 34.538 &$-$01 12 41.29 &1.3, 1.4, 3.5, 7.0 &8.4$\pm$0.3 &1.5$\pm$0.2  &2--5\\
\hline
{\it NE} core &18 47 34.407 &$-$01 12 44.78 &1.3, 1.4, 3.5 &38$\pm$2 &12$\pm$1  &18--38\\
\hline
\end{tabular}

\tiny
$^a$ Wavelengths at which the sources have been detected.  \\
$^b$ Mass estimated at 3.5\,mm using a dust temperature of 50 and 100\,K. \\
$^c$ Mass estimated at 1.4\,mm using a dust opacity of 1.0\,cm$^2$\,g$^{-1}$ and a dust temperature of 50 and 100\,K. \\
\end{table*}

\section{Discussion}

\subsection{No monolithic core}

\begin{figure*}
%\hspace{.5cm}
\centerline{\includegraphics[angle=0,width=16.5cm,angle=0]{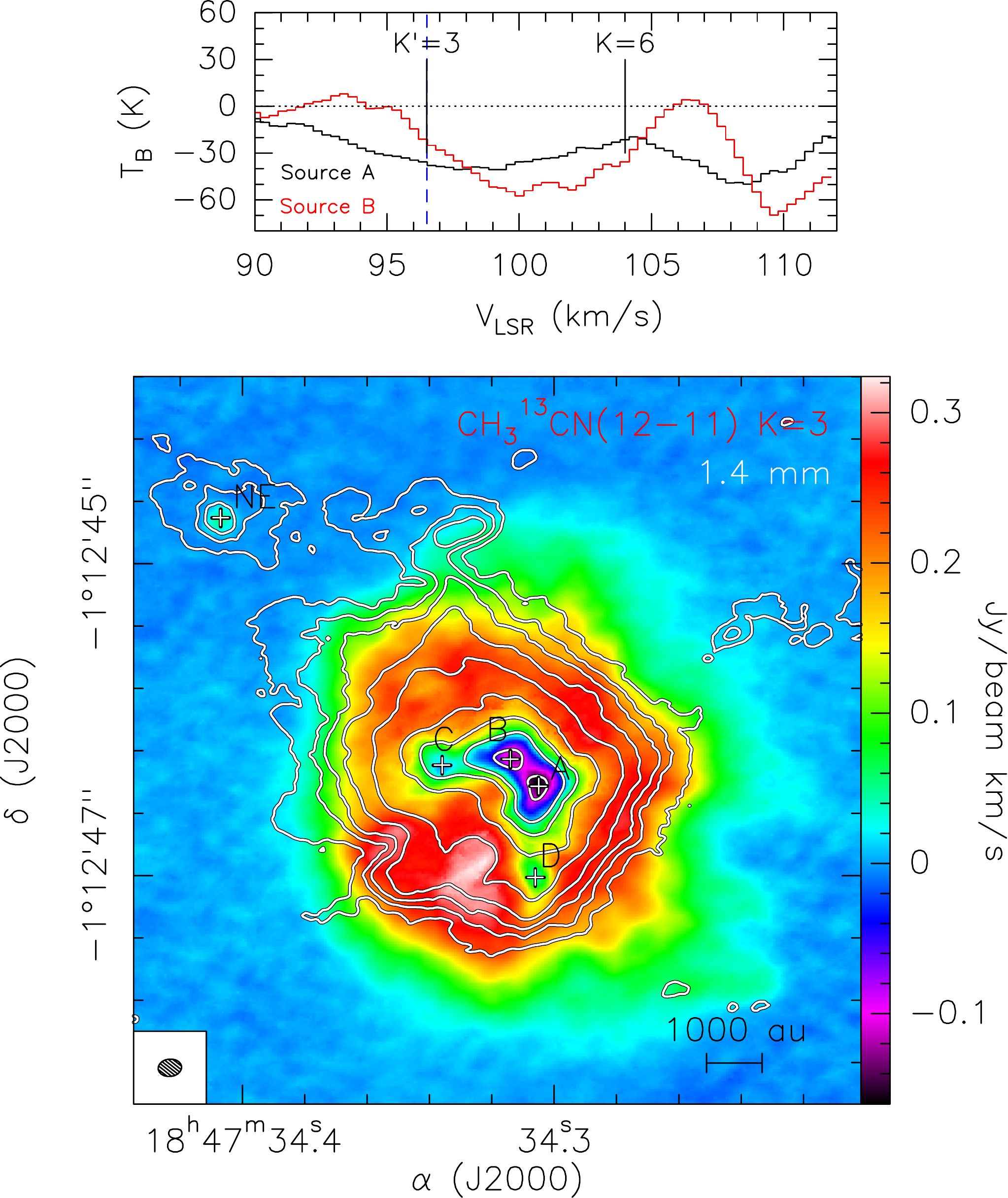}}
\caption{({\it Bottom panel}) Overlay of the ALMA 1.4\,mm continuum emission ({\it white contours}) on the integrated intensity (moment 0) map (colors) of CH$_3^{13}$CN $K = 3$ at an angular resolution of $\sim$0$\farcs1$. The contours are the same as in Fig.~\ref{fig-panels} ({\it right panel}). The synthesized beam is shown in the lower left-hand corner. The white crosses indicate the positions of the {\it NE} core and the four sources embedded in the {\it Main} core. ({\it Top panel}) Spectra of the CH$_3^{13}$CN $K = 3$ and CH$_3$CN $K = 6$ transitions toward the 1.4\,mm continuum emission peak of sources A ({\it black}) and B ({\it red}). The red-shifted absorption is clearly detected in both transitions. The vertical blue dashed line indicates the systemic LSR velocity of 96.5\,km/s.}
\label{fig-ch313cn}
\end{figure*}

The homogeneous, roundish, and seamless morphology of the {\it Main} core in G31 revealed by the $\sim$0$\farcs$2  observations at 1.4\,mm of B18 and 1.3\,mm of Beltr\'an et al.~(\cite{beltran19}) has not been confirmed by our higher angular resolution observations at 1.4\,mm (0$\farcs$1), 3.5\,mm (0$\farcs$075), and 7.0\,mm (0$\farcs$047).  In fact, the {\it Main} core has fragmented into at least four sources within $\sim$1$''$. This confirms what was already suggested by the detection of the free-free continuum sources close to the center (C10), that the apparent homogeneity of the core is due to a combination of the large opacity of the 1.3\,mm and 1.4\,mm dust continuum emission and insufficient angular resolution. Therefore, the new observations have allowed us to discriminate between the two scenarios proposed for the G31 {\it Main} core (see Sect.~\ref{sect-intro}), and suggest that G31 is a dynamically collapsing core undergoing global fragmentation. This conclusion is consistent with the detection of red-shifted absorption toward the sources embedded in the core (e.g., Girart et al.~\cite{girart09}; B18, and Fig.~\ref{fig-ch313cn}), which suggests accretion of material from the common envelope. The analysis of the line emission, and in particular of the kinematics of the embedded sources, will be the subject of a forthcoming paper. 

The brightest sources in the core are A and B and are those located close to its center. These two sources are also the closest to each other in the {\it Main} core, being separated by $\sim$0$\farcs$2 or $\sim$750\,au in projection. Their peak intensities are greater, at some wavelengths by a factor $\gtrsim$5, than those of sources C and D.  Sources A and B have very similar SEDs (see Fig.~\ref{fig-sed}), with source A being slightly stronger at 3.5\,mm, 7\,mm, and 1.3\,cm. Similarities between the two sources can also be found in their sizes as shown in Table~\ref{table-cont}. The diameter of the sources at 3.5\,mm, where the emission is tracing mostly thermal dust emission, is $\sim$800\,au. Note that the size of the sources at the highest angular resolution achieved with our observations (0$\farcs$047 at 7\,mm) is only $\sim$200\,au. However, as discussed in  Sect.~\ref{sect-free}, because the emission at 7\,mm is highly contaminated by free-free emission, this should be considered the size of the thermal radio jets associated with the sources and not the size of the dusty cores. The gas masses and densities of sources A and B are also similar (Table~\ref{table-mass}). The masses, which are $\sim$15\,$M_\odot$, are consistent with the sources being associated with high-mass (proto)stars. This is confirmed by their high mass surface densities ($\sim$270\,g\,cm$^{-2}$; see Table~\ref{table-mass}) and the detection of infalling gas toward them.  The similar properties and the relative small separation of sources A and B might indicate a common origin. The sources could be members of a binary system with a gas mass ratio close to 1. Photometric and spectroscopic observations of more evolved OB stars indicate a high rate ($>$50\%) of close binaries (Chini et al.~\cite{chini13}), with similar masses of the binary members. Hydrodynamical simulations of star-forming clusters predict a high fraction of binaries with similar masses for systems with separations $<$10~au (e.g, Lund \& Bonnell~\cite{lund18};  Clarke~\cite{clarke07} and references therein), much smaller than that between source A and B. On the other hand, 3D radiation-hydrodynamic simulations of high-mass star formation by Krumholz et al.~(\cite{krumholz09}) predict the formation of massive binary systems at the center of the collapsing cores with separations between the binary members similar to that of sources A and B, and with a final stellar mass ratio of $\sim$0.7. Therefore, these observational and theoretical works suggest that massive binary systems have mass ratios close to 1, while our results suggest that this mass ratio might be set already at the very early evolutionary stages of high-mass star formation. Note that this conclusion holds if not only the gas masses but also the (proto)stellar masses of the objects embedded in sources A and B are similar, as suggested by the simulations of Bate~(\cite{bate00}). This study predicts that, regardless of their initial mass ratio, massive binaries will rapidly evolve toward a mass ratio of 1. In order to test the hypothesis of binarity in G31, one should analyze the line emission toward the sources to search for dynamical features that link them or reveal infalling material from a circumbinary disk.

\begin{figure*}
%\hspace{.5cm}
\centerline{\includegraphics[angle=0,width=17cm,angle=0]{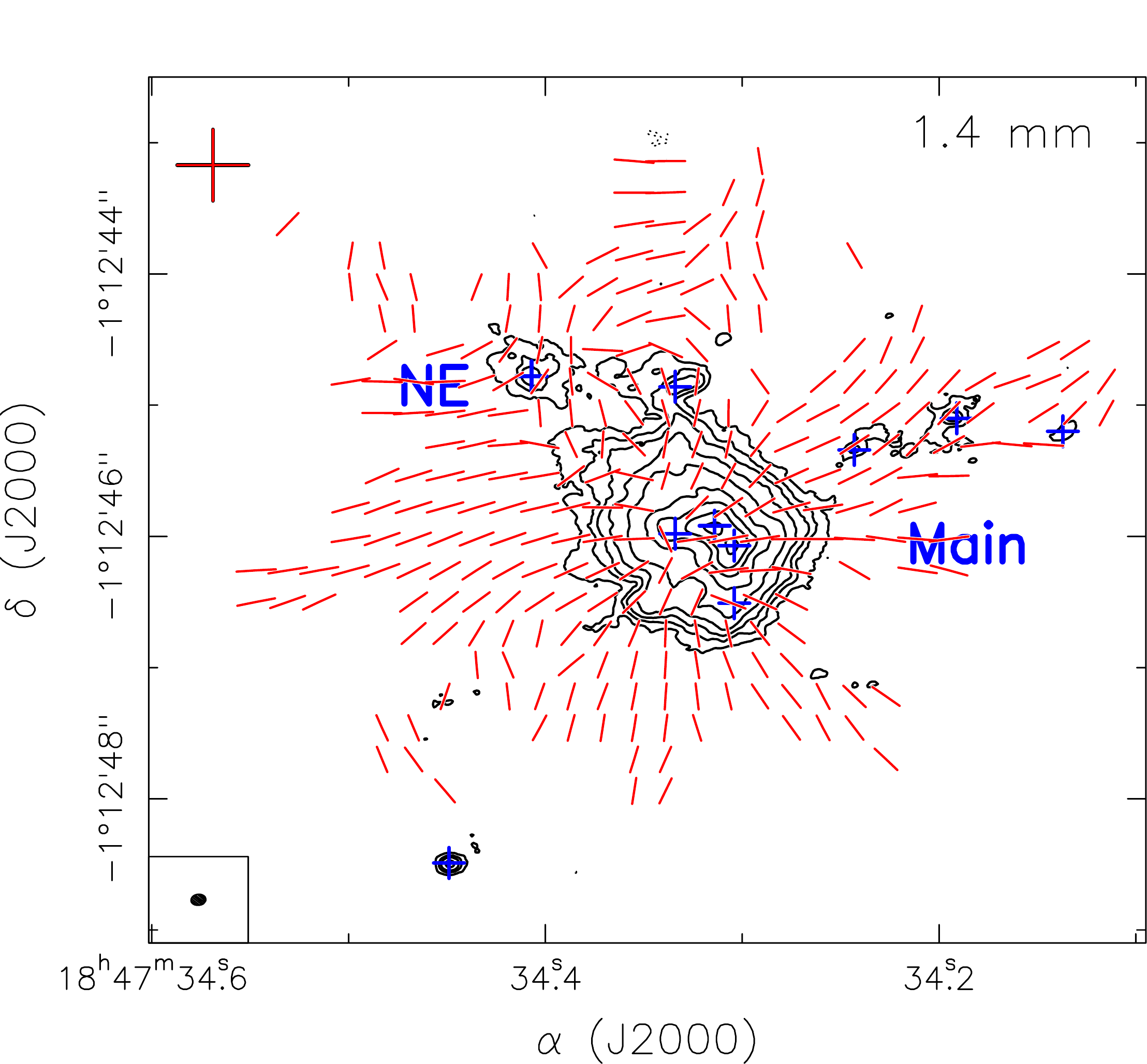}}
\caption{Magnetic field segments ({red lines}) toward the G31 HMC, obtained by rotating 90$^\circ$ the polarization segments at 1.3\,mm from Beltr\'an et al.~(\cite{beltran19}), overlapped on the ALMA 1.4\,mm continuum emission map ({contours}). The segments are sampled every ten pixels. The contours are the same as in Fig.~\ref{fig-allcore}a.  Negative contours are dotted black. The synthesized beam is shown in the lower left-hand corner. The blue and red crosses indicate the position of the millimeter and/or centimeter sources in the region.}
\label{fig-magnetic}
\end{figure*}

The other two sources embedded in the {\it Main} core, sources C and D, are also massive, as indicated by their gas masses of $\sim$26\,$M_\odot$. These two sources are more extended, especially at 3.5\,mm, and colder than sources A and B (Table~\ref{table-mass}), which suggests that they could be in a slightly earlier evolutionary phase. The enhanced surface densities and top-heavy mass function of the fragments in the {\it Main} core are similar to the properties predicted by the hub-filament system model (Kumar et al.~\cite{kumar20}), because hubs have amplified densities and low virial parameters.

\begin{figure*}
\centerline{\includegraphics[angle=0,width=17cm,angle=0]{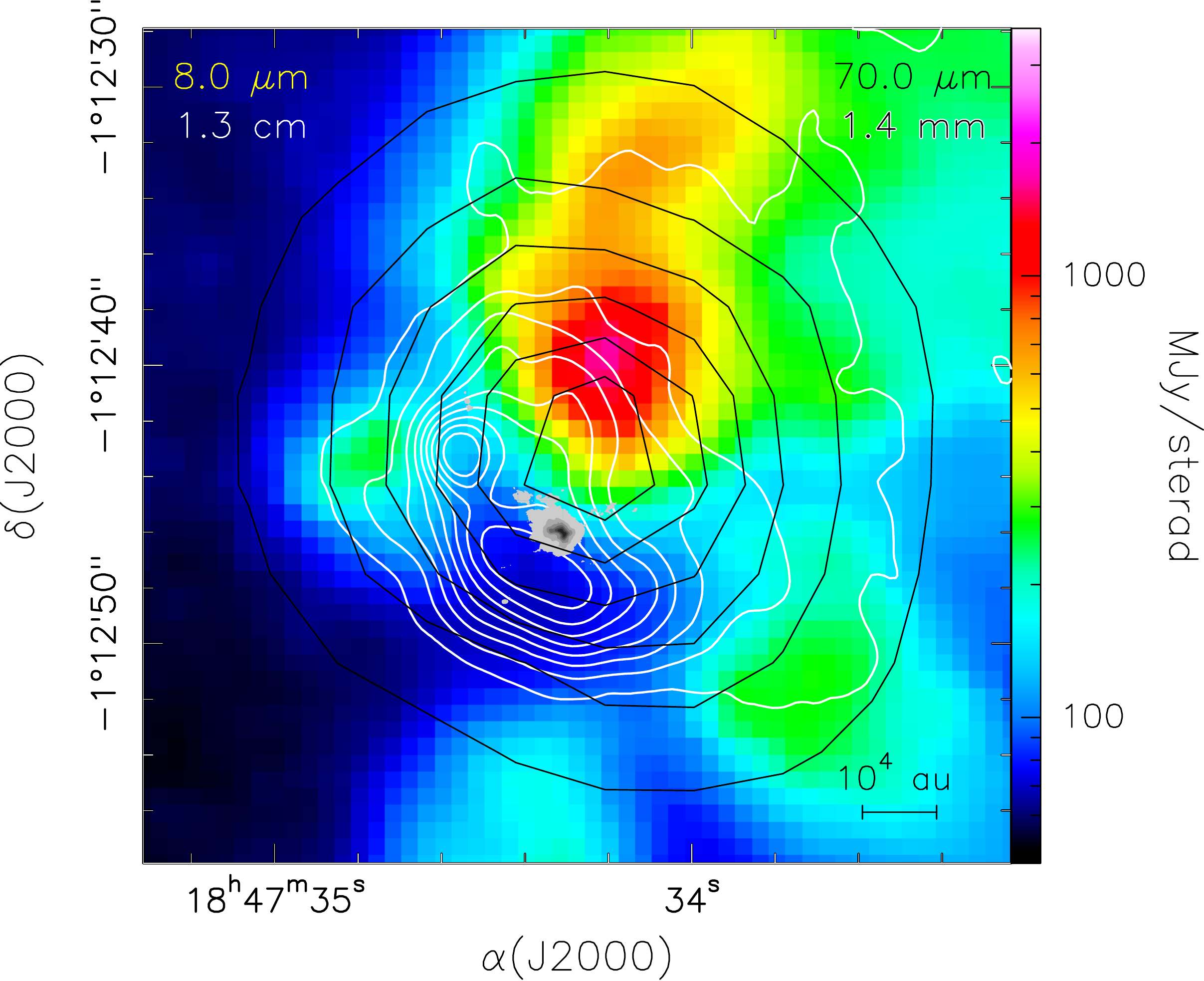}}
\caption{Overlay of the ALMA 1.4\,mm continuum emission ({\it greyscale}), the VLA combined C \& D Array continuum emission at 1.3\,cm ({\it white contours}) from Cesaroni et al.~(\cite{cesa94}), and the $Herschel$ Hi-GAL 70\,$\mu$m emission ({\it black contours}) on the  $Spitzer$ GLIMPSE 8.0\,$\mu$m emission ({\it colors}). White contours range from 4\,\mjy\ to 60\,\mjy\ by steps of 7\,\mjy. Black contours range from 2000\,MJy\,sterad$^{-1}$ to 152\,000\,MJy\,sterad$^{-1}$ by steps of 30\,000\,MJy\,sterad$^{-1}$.}
\label{fig-irac}
\end{figure*}

\subsection{Fragmentation and magnetic field}

From the analysis of the 1.3\,mm dust polarization maps (see Fig.~\ref{fig-magnetic}), the G31 HMC represents one of the clearest examples, to date, of an hourglass-shaped magnetic field morphology in the high-mass regime (Girart et al.~\cite{girart09}), with the magnetic field, oriented southeast to northwest, maintaining its morphology 
down to scales of $<$1000\,au (Beltr\'an et al.~\cite{beltran19}). 
The central region of the  core has been found to be moderately magnetically supercritical, with a mass-to-flux ratio $\lambda$ about $1.4$--$2.2$ above the critical value and an estimated  strength of the magnetic field  in the range $8$--$13$\,mG (Beltr\'an et al.~\cite{beltran19}). The alignment of sources A, B, and C, D 
along a direction NE-SW roughly perpendicular to the axis of the magnetic field 
%inferred by Beltr\'an et al.~(\cite{beltran19}),
also supports a scenario of magnetically-regulated fragmentation. The position of the four sources 
in the {\it Main} core, found close to the ``waist'' of the hourglass field, suggests the possibility that they formed from the fragmentation of a massive pseudo-disk created by the magnetic focusing of the collapsing gas (Galli \& Shu~\cite{galli93}). Given the inclination of the magnetic axis of the hourglass with respect to the plane of the sky of 
$\sim 44^\circ$ (Beltr\'an et al.~\cite{beltran19}), the relative positions of the four sources are consistent with formation in a large, inclined, flattened dense structure at the center of the G31 {\it Main} core. 

Magneto-hydrodynamical (MHD) simulations of the collapse 
of rotating, magnetized clouds (e.g., Hosking \& Whitworth~\cite{hosking04}; 
Machida et al.~\cite{machida05a}, \cite{machida05b};
Price \& Bate~\cite{price08}; Peters et al.~\cite{peters11}) show that fragmentation is generally inhibited by dynamically important levels of magnetization: for moderate to high values of the mass-to-flux ratio $\lambda$ ($\sim 2$--$14$ times the critical value), the number of fragments formed during collapse is roughly reduced
by half, with respect to the pure hydrodynamical case, whilst for values of mass-to-flux ratio closer to the critical value,
fragmentation is completely suppressed and only one 
single object is formed (Hennebelle et al.~\cite{hennebelle11}; Commer\c con et al.~\cite{commercon11}). However, the exact outcome of the  process of fragmentation also depends on other factors, such as the density profile of the collapsing core/cloud (e.g., Girichidis et al.~\cite{girichidis11}), radiative and  mechanical (jet/outflow) feedback from the newly born stars,  turbulence, and the amount of rotation of the initial configuration. In fact, a notable feature of the G31 {\it Main} core is its relatively high angular rotation, of the order of $\Omega\sim 5\times 10^{-12}$~s$^{-1}$, if the velocity gradient of $\sim 160$~km~s$^{-1}$~pc$^{-1}$ observed by B18 in the \MCNII(12--11) $K=2$ transition is interpreted as solid-body rotation. With a mean density of $\sim 4\times 10^7$~cm$^{-3}$,  estimated from the mass of 70\,$M_\odot$ (Cesaroni et al.~\cite{cesa19}) and a radius of $\sim$1$\farcs$076 or $\sim$4000\,au (B18), the characteristic 
free-fall time of the {\it Main} core is  $t_{\rm ff}\sim 10^4$~yr, which implies $\Omega\, t_{\rm ff} \sim 1$. According to the MHD collapse simulations by Machida et 
al.~(\cite{machida05a}, \cite{machida05b}), during the collapse of a model cloud  with levels of magnetization and rotation similar to  those of the G31 {\it Main} core (their  model D: $\lambda = 1.4$, $\Omega\, t_{\rm ff}=3$) any initial non-axisymmetric perturbation is  strongly amplified, and the dense core 
of size $\sim$100\,au formed at the end of the isothermal phase is prone to bar instability and rotationally-induced fragmentation. This is similar to the ionization-MHD simulations of Peters et al.~(\cite{peters11}). To establish whether this scenario can be applied to the case of G31, the enigmatic interplay between rotation and magnetic fields in the evolution of the G31 {\it Main} core (Girart et al.~\cite{girart09}; B18) deserves consideration by further studies.

Further indications of a magnetically-controlled 
fragmentation come from considering the Jeans mass
of the {\it Main} core. Assuming that the temperature of 
the {\it Main} core before the formation of the embedded sources was $\sim 20$~K, as suggested
by observations of high-mass starless
cores (e.g., Fontani et al.~\cite{fontani15}), 
and assuming an initial uniform density similar to the mean density of the {\it Main} core, $\sim$4$\times10^7$\,cm$^{-3}$,
%the thermal Jeans length of the Main core is $\sim 500$~au
the initial Jeans mass in the {\it Main} core would have been $\sim 0.1$~$M_\odot$. The assumption of a density at the onset of fragmentation similar to the current one is based on the fact the the Jeans mass has a much stronger dependence on temperature than on density, and that the initial density of the {\it Main} core should not have been much lower than the current density. Moreover, the current temperature is the result of the heating of the gas by the newly formed stars, and the gas heating has occurred on a time scale much shorter than the evolutionary time scale of the {\it Main} core that characterizes the density increase. Therefore, the density variation during the heating process should be negligible. 
%a factor of $(20/250)^{3/2}\approx 0.02$ smaller than the present-day value, obtained for the mean temperature of the core of 250\,K, estimated from Eq.~(1) of B18. 
Clearly, a cloud with these properties would  
have produced a large 
number of fragments of small mass, unlike what observed in G31. This region appears to have turned into a power-law density profile by global gravitational
contraction, dragging in the magnetic field initially present in the cloud, and producing in this process a relatively  small number of massive fragments, mostly concentrated in the central part. Although highly speculative, these considerations support the view that the G31 {\it Main} core has undergone a phase of magnetically-regulated fragmentation characterized by a reduced fragmentation efficiency, eventually leading to the formation of a small number of relatively massive dense cores, consistent with MHD simulations (e.g., Machida et al.~\cite{machida05a}, \cite{machida05b}; Commer\c con et al.~\cite{commercon11}; Peters et al.~\cite{peters11}; Myers et al.~\cite{myers13}). 

Whether or not the process of fragmentation is now completed in the G31 {\it Main} core is difficult to assess. According to Krumholz \& McKee~(\cite{krumholz08}), radiative feedback effects suppress further fragmentation in cores with a surface density larger  than $\sim$1\,g\,cm$^{-2}$, as the radiation from the 
embedded protostars heats the gas providing efficient
contrast to further local collapse. 
%The present-day Jeans mass of the Main core of
%for a mean temperature of 250\,K and a mean number density of $\sim$4$\times10^7$\,cm$^{-3}$, 
%the Jeans length of the Main core
%is $\sim 3500$~au, comparable with its radius of $\sim 4000$~au.
%$\sim$45\,$M_\odot$, comparable to the estimated
%mass of the Main core. 
%As noted in Sect.~4.1, the surface density of the %Main core is quite high, $\sim$12\,g\,cm$^{-2}$.
For the G31 {\it Main} core, the surface density estimated for a mass of 70\,$M_\odot$ and a radius of $\sim$4000\,au is quite high, $\sim$12\,g\,cm$^{-2}$. This suggests that
thermal (and possibly also mechanical) feedback from the embedded sources may have halted the fragmentation of the {\it Main} core.

%Similarly, the two prominent sources A and B in the G31 %Main core appear to be 
%close to a state of approximate balance between thermal %and 
%gravitational energy: with the parameters listed in %Table~3, 
%their associated Jeans masses are $\sim 6$~$M_\odot$, 
%very close to their actual estimated masses; sources C %and D, 
%on the other hand, have masses roughly equal to half %their associated
%Jeans masses of $\sim 5$ and $\sim 3$~$M_\odot$, %respectively.

\bigskip
\bigskip

\subsection{The global picture of the G31.41+0.31 high-mass star-forming region} 

Figure~\ref{fig-irac} shows a large-scale image of the G31.41+0.31 high-mass star-forming region at different wavelengths: infrared, millimeter, and centimeter. The white contours show the continuum emission at 1.3\,cm observed with the C \& D configurations of the VLA by Cesaroni et al.~(\cite{cesa94}). The centimeter emission, which is dominated by the free-free emission of the UC \HII region, is very extended and has a cometary shape with two peaks located close to the ``head''. The strongest and most compact peak is located to the east and is likely tracing the position of the O6 star ionizing the UC \HII region (C10), while the secondary and more elongated peak is located to the south close to the position of the G31 HMC. This secondary peak could indicate a zone where the UV photons escaping from the central part of the UC \HII region ionize the dense material associated with the HMC, creating a sort of bright-rimmed ridge.  The ``tail'' of the cometary free-free emission points northward, where the region is probably less dense. This is further supported by the fact that the ``tail''  of the free-free emission coincides with the peak of the infrared emission at 8.0\,$\mu$m from {\it Spitzer}, which suggests that the photons from the stars embedded in the G31 HMC escape through the lower-density region. The peak of the infrared emission at 70\,$\mu$m, as seen by {\it Herschel}, is also located northward of the G31 HMC, but closer to it than the peak of the 8\,$\mu$m emission, consistent with the lower dust opacity at 70\,$\mu$m. This seems to confirm that the heating sources are indeed embedded inside the HMC. 

%The \HII region is probably the most evolved source in the star-forming region. If we zoom in toward the position of the HMC, we should find  

Zooming in toward the position of the HMC, we find a small (proto)cluster of millimeter sources, probably in an earlier evolutionary stage with respect to the UC \HII region. This small (proto)cluster consists of twelve young stellar objects, mostly concentrated in a small region of $\sim$5$''$ around the central position of the {\it Main} core, as seen in Fig.~\ref{fig-allcore}. Besides the four sources embedded in the {\it Main} core (A, B, C, and D), there are six additional sources (\#1, 2, 3, 4, 5, and the {\it NE} core), very close to it and located in what appears as  streams/filaments of material pointing to the {\it Main} core. This would suggest a scenario similar to that described by the  ``competitive accretion'' model of Bonnell \& Bate~(\cite{bonnell06}), that predicts that a molecular cloud initially fragments in low-mass cores of Jeans mass that form stars that compete to accrete mass from the common gas reservoir.  Protostars located near the center of the gravitational potential accrete at a higher rate because of a stronger gravitational pull, and thus experience a faster mass growth. It would also be consistent with the  ``fragmentation-induced starvation'' scenario of Peters et al.~(\cite{peters10}). According to this model, the dense accretion flow onto the central massive object becomes gravitationally unstable leading to  fragmentation and formation of secondary objects in the flow that starve the central star of accreting material.  In G31, at the center of the cloud we find a small cluster of four massive young stellar objects, sources A, B, C, and D, still actively accreting, as demonstrated by the detection of red-shifted absorption and molecular outflows associated with them. 

From an observational point of view, the large-scale scenario in the G31 star-forming region resembles that observed in some of the densest and most massive HMCs forming protoclusters: e.g., Sgr\,B2(N) (S\'anchez-Monge et al.~\cite{sanchez-monge17}; Schw\"orer et al.~\cite{schwoerer19}) and W51 (Goddi et al.~\cite{goddi20}), where complex dusty streams and filaments converging onto the central compact sources have been detected. These filaments are dense and relatively small, with spatial scales of $\sim$0.1\,pc for Sgr\,B2(N) (S\'anchez-Monge et al.~\cite{sanchez-monge17}) and $<$2000\,au for W51 (Goddi et al.~\cite{goddi20}). Schw\"orer et al.~(\cite{schwoerer19}) have analyzed the velocity field along the filaments in Sgr\,B2(N) and found velocity gradients consistent with accretion to the center of the cores, similarly to what Peretto et al.~(\cite{peretto13}) have found in much younger (infrared-dark) high-mass star-forming cores. For W51, Goddi et al.~(\cite{goddi20}) have not been able to 
measure the gas kinematics along the dusty streams because of the large opacity of the 1.3\,mm dust continuum emission. In G31, we have searched for velocity gradients along the dust streams in which the sources \#1, 2, 3, 4, 5, and the {\it NE} core seem to be located, to identify kinematic signatures of accretion of material onto the central region from the larger-scale cloud. Unfortunately, the current high-angular resolution line data do not have enough sensitivity to allow us to study the velocity field, and we cannot completely confirm that such scenario is in place in G31.

%Not the most massive and evolved star would be a bit isolated. 
%Nor the more evolved.

\section{Conclusions}

We carried out ALMA observations at 1.4\,mm and 3.5\,mm of the high-mass star-forming region G31 
at angular resolutions of $\sim$0$\farcs$1 ($\sim$375\,au) and $\sim$0$\farcs$075 ($\sim$280\,au), respectively. The goal of these observations was to establish whether the {\it Main} core in G31 is as monolithic as it appears at $\sim$1\,mm and $\sim$0$\farcs$2 angular resolution (B18, Beltr\'an et al.~\cite{beltran19})  or whether its homogeneous appearance is due to a combination of both large dust opacity and insufficient angular resolution. We also carried out VLA observations at 7\,mm and
1.3\,cm at $\sim$0$\farcs$05 ($\sim$190\,au) and $\sim$0$\farcs$07 ($\sim$260\,au) angular resolutions, respectively, to better study the nature of the free-free continuum sources previously detected by C10.

The millimeter continuum emission of the {\it Main} core has been clearly resolved into at least four sources, A, B, C, D, within $\sim$1$''$. Sources A, B, and D have also been detected at centimeter wavelengths. In all three cases, the free-free emission is likely associated with thermal radio jets. Besides these four sources, eight more have been detected in the region within 12$''$, including the \UC\ region located $\sim$5$''$ northeast from the {\it Main} core. The \UC\ has been detected at 1.3\,cm, 7\,mm, and 3.5\,mm and shows a spherical and compact morphology, with a diameter of $\sim$1$''$ or 3750\,au (0.018\,pc) at the distance of the region.  

The masses of the four sources embedded in the {\it Main} core, estimated at 1.4\,mm, range from $\sim$15 to $\sim$26\,$M_\odot$ and their number densities are several 10$^9$\,cm$^{-3}$. The deconvolved radii of the dust emission of the sources, estimated at 3.5\,mm,  range from $\sim$400 to 500\,au, while the radii of the free-free emission, as estimated at 7\,mm or 1.3\,cm, range from $\sim$100 to 175\,au. The mass surface densities of all the sources are very high, with values 250--350\,g\,cm$^{-2}$. 

Sources A and B, separated by $\sim$750\,au, dominate the flux density of the {\it Main} core and are those located close to its center. The millimeter emission peak coincides with the centimeter emission peak in both sources, proving that the free-free emission from sources A and B is tracing two distinct (jets from) young stellar objects and not the lobes of the same radio jet. Sources A and B have similar SEDs, masses, and sizes, and both are undergoing collapse, as suggested by the detection of red-shifted absorption toward them (see Fig.~\ref{fig-ch313cn}). Altogether, this indicates that the sources are associated with
high-mass (proto)stars. This is confirmed by the high mass surface densities. Sources C and D, which are massive as well, have a more extended morphology and lower temperatures and could be high-mass young stellar objects in an earlier evolutionary phase. Source D is associated with centimeter emission likely arising from a thermal radio jet.

%This scenario resembles the collapse of a massive %prestellar core described by the 3D %radiation-hydrodynamic simulations of Krumholz et %al.~(\cite{krumholz09}), in which a massive binary %system forms at the center of the core, with each of %the two accreting massive protostars having its own %disk. 

The observations have confirmed that the collapsing {\it Main} core in G31 has undergone fragmentation and therefore that the previously assessed monolithic appearance of G31 is a consequence of both insufficient angular resolution and dust opacity of the core. %The separation of the sources is consistent with the thermal Jeans length of $\sim$630\,au. 
%The initial Jeans mass in the central region would have been of $\sim$1\,M$_\odot$, 
%much smaller than the mass of the core, $\sim$70\,$M_\odot$. Therefore, t
The low level of fragmentation suggests that the G31 {\it Main} core has undergone a phase of magnetically-regulated evolution characterized by a reduced fragmentation efficiency, eventually leading to the formation of a small number of relatively massive dense cores, as found by theoretical models and simulations.

\begin{acknowledgements}

We thank the anonymous referee for their
useful comments that have improved the manuscript. This paper makes use of the following ALMA data: ADS/JAO.ALMA\#2013.1.00489.S, ADS/JAO.ALMA\#2016.0.00223.S, and ADS/JAO.ALMA\#2018.0.00252.S. ALMA is a partnership of ESO (representing its member states), NSF (USA) and NINS (Japan), together with NRC (Canada) and NSC and ASIAA (Taiwan), in cooperation with the Republic of Chile. The Joint ALMA Observatory is operated by ESO, AUI/NRAO and NAOJ.   The National Radio
Astronomy Observatory is a facility of the National Science Foundation operated under cooperative agreement by Associated Universities, Inc. V.M.R.\ acknowledges support from the Comunidad de Madrid through the Atracci\'on de Talento Investigador Modalidad 1 (Doctores con experiencia) Grant (COOL: Cosmic Origins Of Life; 2019-T1/TIC-15379), and from the  European Union's Horizon 2020 research and innovation programme under the Marie Sk\l{}odowska-Curie grant agreement No 664931. H..\ acknowledges support from the European Research Council under the Horizon 2020 Framework Program via the ERC Consolidator Grant CSF-648405. H.B.\ also acknowledges support from the Deutsche Forschungsgemeinschaft in the Collaborative Research Center (SFB881) ``The Milky Way System'' (subproject B1). C.G.\  acknowledges support by the ERC Synergy Grant ``BlackHoleCam: Imaging the Event Horizon of Black Holes'' (Grant 610058). R.K.\ acknowledges financial support via the Emmy Noether Research Group on Accretion Flows and Feedback in Realistic Models of Massive Star Formation funded by the German Research Foundation (DFG) under grant no.\ KU 2849/3-1 and KU 2849/3-2. M.S.N.K.\ acknowledges the support from FCT - Fundac\~{a}o para a Ci\^{e}ncia e a Tecnologia through Investigador contracts and exploratory project (IF/00956/2015/CP1273/CT0002) 
\end{acknowledgements}

\end{document}